\documentclass[fleqn,usenatbib]{mnras}

\usepackage{newtxtext,newtxmath}

\usepackage[T1]{fontenc}

\DeclareRobustCommand{\VAN}[3]{#2}
\let\VANthebibliography\thebibliography
\def\thebibliography{\DeclareRobustCommand{\VAN}[3]{##3}\VANthebibliography}

\usepackage{xspace}
\usepackage{enumitem}
\usepackage{todonotes}
\usepackage{orcidlink}
\usepackage{gensymb}
\usepackage{booktabs} 

\usepackage{multirow} 
\usepackage{graphicx} 
\usepackage{array} 
\usepackage{tabularx,ragged2e,caption, relsize}

\graphicspath{{./figs/}}

\newcommand{\src}{{SDSSJ2306}} 
\newcommand{\arsco}{{AR\,Sco}}
\newcommand{\wdp}{{J1912}}
\newcommand{\srclong}{{SDSS~J230641.47$+$244055.8}}

\newcommand{\oklahoma}{7}
\newcommand{\colorado}{8}
\newcommand{\sheffield}{4}
\newcommand{\iac}{5}
\newcommand{\cork}{6}
\newcommand{\hamburg}{9}

\newcommand{\Msun}{\mbox{$\mathrm{M}_\odot$}}

\title[\src\ and the Blueprint of WD pulsars]{A Sibling of AR Scorpii: SDSS\,J230641.47$+$244055.8 and the Observational Blueprint of White Dwarf Pulsars}

\author[N. Castro Segura et al.]{N.~Castro~Segura$^{\orcidlink{0000-0002-5870-0443}}$\,$^{1}$\thanks{E-mail: Noel.Castro-Segura@Warwick.ac.uk},
I.~Pelisoli$^{\orcidlink{0000-0003-4615-6556}}$\,$^{1}$,
B.~T.~G\"ansicke$^{\orcidlink{0000-0002-2761-3005}}$\,$^{1}$,
D.~L.~Coppejans$^{\orcidlink{0000-0001-5126-6237}}$\,$^{1}$,
D.~Steeghs$^{\orcidlink{0000-0003-0771-4746}}$\,$^{1}$,
\newauthor
A. Aungwerojwit$^{\orcidlink{0000-0001-5686-5650}}$\,$^{2}$,
K.~Inight$^{\orcidlink{0000-0002-2200-2416}}$\,$^{1}$,
A.~Romero$^{\orcidlink{0000-0002-0797-0507}}$\,$^{3}$,
A. Sahu$^{\orcidlink{0009-0007-6825-3230}}$\,$^{1}$,
V.~S. Dhillon$^{\orcidlink{0000-0003-4236-9642}}$\,$^{\sheffield,\iac}$,
J. Munday$^{\orcidlink{0000-0002-1872-5398}}$\,$^{1}$,
\newauthor
S.~G. Parsons$^{\orcidlink{0000-0002-2695-2654}}$\,$^{\sheffield}$,
M. R. Kennedy$^{\orcidlink{0000-0001-6894-6044}}$\,$^{\cork}$,
M.~J. Green$^{\orcidlink{0000-0002-0948-4801}}$\,$^{\oklahoma,\colorado}$
A.~J. Brown$^{\orcidlink{0000-0002-3316-7240}}$\,$^{\hamburg}$,
M.~J. Dyer$^{\orcidlink{0000-0003-3665-5482}}$\,$^{\sheffield}$,
E. Pike$^{\sheffield}$,
\newauthor
J.~A. Garbutt$^{\sheffield}$,
D. Jarvis$^{\sheffield}$,
P. Kerry$^{\sheffield}$,
S.~P. Littlefair$^{\orcidlink{0000-0001-7221-855X}}$\,$^{\sheffield}$,
J. McCormac$^{1}$,
D.~I. Sahman$^{\sheffield}$,
\newauthor
D.~A.~H.
Buckley$^{\orcidlink{0000-0002-7004-9956}}$\,
$^{10,11}$
\newauthor
\emph{\normalsize Affiliations are listed at the end of the paper}
}

\date{Accepted XXX. Received YYY; in original form ZZZ}

\pubyear{2025}

\begin{document}
\label{firstpage}
\pagerange{\pageref{firstpage}--\pageref{lastpage}}
\maketitle

\begin{abstract}
Radio pulsating white dwarf (WD) systems, known as WD pulsars, 
are non-accreting binary systems where the rapidly spinning WD interacts with a low-mass companion producing pulsed non-thermal emission that can be observed across the entire electromagnetic spectrum. 
Only two such systems are known: \arsco\ and eRASSU\,J191213.9$-$441044. Here we present the discovery of a third WD pulsar, \srclong. The optical spectrum is dominated by molecular bands from an M-dwarf companion, with additional narrow emission lines from the Balmer series and He~{\sc i}. 
The long-term optical light-curve folded on its orbital period ($P_\mathrm{orb} = 3.49$ h) exhibits large scatter (roughly 10 per cent). High-cadence photometry reveals a short period signal, which we interpret to be the spin period of the WD primary ($P_\mathrm{spin} \simeq 92$\,s). The WD spin period is slightly shorter than that of \arsco\ ($\rm \sim 117$\,s), the WD pulsar prototype. 
Time-resolved spectroscopy reveals emission from the irradiated companion and Na\,{\sc i} absorption lines approximately tracing its centre of mass, which yields a binary mass function of $f(M) \simeq 0.2\,{\rm M_\odot}$. The H$\alpha$ emission includes a low-amplitude broad component, resembling the energetic emission line flashes seen in AR\,Sco.
Using spectral templates, we classify the companion to be most likely a $\rm M4.0\pm 0.5$ star with $T_\mathrm{\rm eff} \approx 3300$\,K. Modelling the stellar contribution constrains the secondary mass ($0.19\,\Msun\lesssim M_2\lesssim 0.28\,\Msun$), system distance ($\simeq1.25\,{\rm kpc}$), and inclination ($i \simeq 45-50^\circ$).
We discuss the proposed evolutionary scenarios and summarize the observational properties of all three known WD pulsars, establishing a benchmark for identifying and classifying future members of this emerging class.

\end{abstract}




\section{Introduction}

White dwarf (WD) pulsars are non-accreting binary systems composed of a nearly Roche lobe filling late-type main sequence secondary star with a rapidly spinning primary WD. The luminosity is higher than the stellar contributions alone, and is thought to be powered by the spin down of the WD. 
To date, only two such systems are known — \arsco\ and eRASSU J191213.9$-$441044 — making them an exceptionally rare class of compact binaries.
These systems are characterized by exhibiting strong pulsed emission across the entire electromagnetic spectrum (from radio to X-rays) associated with the spin period of the WD \citep{Marsh2016Natur.537..374M,AR_Sco_Stanway:2018A&A...611A..66S,Takata_ARSco_XMM:2018ApJ...853..106T, Pelisoli2023NatAs...7..931P,Schwope_J1912:2023A&A...674L...9S}. 
One of the most distinctive observational features of WD pulsars arises from their optical light curve. In their monochromatic lightcurves, white dwarf pulsars exhibit an orbital modulation with a single maximum per orbital cycle, which suggests the presence of a hot component \citep[e.g.][]{NNSer_UCAM_Brinkworth:2006MNRAS.365..287B} that is not directly detected in their optical spectra.
Unlike accreting systems, whose short-term variability is dominated by stochastic ``red'' noise \citep[e.g. ][]{Scaringi_rms-flux:2015SciA....1E0686S}, the observed light curves of WD pulsars exhibit highly coherent pulses, signalling the absence of accretion onto the WD \citep{Pelisoli_ARSco_long-term:2022MNRAS.516.5052P}. The lack of significant mass transfer between the binary components is also supported by their optical/near-IR spectroscopic properties and low X-ray luminosities. In particular, their observed spectra are dominated by the molecular bands typical of a relatively cold M-dwarf star with very narrow emission lines originating from its irradiated face \citep[e.g. ][]{Garnavich_ARSco_time_resolved_sp:2019ApJ...872...67G}.

At optical wavelengths, the pulsed emission can outshine the stellar components and exhibits a strong linear polarization, up to 40 per cent with up to a 90 per cent pulse fraction in the WD pulsar prototype, \arsco. These polarized components show periodicities consistent with both the spin period and the beat of the spin and orbital periods, suggesting that the emission is driven by the interaction of the WD's magnetosphere with the atmosphere of the stellar companion \citep[][]{Buckley_ARSco:2017NatAs...1E..29B}. This is consistent with the location of energetic emission line flashes observed in \arsco\ \citep[][]{Garnavich_ARSco_time_resolved_sp:2019ApJ...872...67G}, however the mechanism behind the pulsing behaviour is still debated \citep[e.g. ][]{Geng_ARSco:2016ApJ...831L..10G,Katz_ARSco_reconnection:2017ApJ...835..150K,Takata_ARSco_XMM:2018ApJ...853..106T,Potter_Buckley_ARSco:2018MNRAS.481.2384P,duPlessis2022MNRAS.510.2998D,Barrett_ARSco_SED:2025arXiv250506468B}.

Regardless of the details on how the pulsed emission is generated, WD pulsars are instrumental in our understanding of the formation and evolution of magnetic WDs. WD pulsars may be the missing link between the two main classes of accreting magnetic WDs, which represent a large fraction of interacting WD binaries \citep[][]{Pala_volumelimitedsample:2020MNRAS.494.3799P,Rodriguez_pop_xr:2025PASP..137a4201R}. These are known as polars and intermediate polars \citep[a.k.a. AM Her and DQ Her stars respectively; e.g.][]{Warner1995cvs..book.....W}, and they are distinguished by having spin and orbital period synchronised (polars) or asynchronous (intermediate polars), as a consequence of their relative difference in magnetic field strength with the orbital period, which can magnetically lock the orbit with the WD rotation (with typical magnetic fields strengths $\mathrm{B}> 10$\,MG for polars, and below that for intermediate polars). 
\citet{Schreiber_B_in_WDs:2021NatAs...5..648S} proposed an evolutionary scenario where the WDs in accreting binaries initially have weak magnetic fields, allowing for efficient accretion that spins up the WD primary. As they cool down and their cores starts to crystallize, the rapid rotation may trigger a dynamo, generating a magnetic field \citep[e.g. ][]{Isern_B_in_WDs:2017ApJ...836L..28I}. The system then appears as an intermediate polar. As the field strengthens, magnetic coupling with the companion slows the WD's spin via transfer of angular momentum from the WD to the orbit, potentially detaching the system and producing a WD pulsar stage. Continued angular momentum loss will eventually bring the system into contact, thus restarting the mass transfer and evolving the system into a polar. 
It is important to note that the efficiency of the crystallisation dynamo originally incorporated in this model is under discussion, with recent works suggesting that it would not generate a strong enough magnetic field \citep{Castro-Tapia_crystallization:2024ApJ...975...63C, Camisassa2024}. In any case, there is observational evidence that white dwarfs become more magnetic as they age \citep[e.g.][]{Bagnulo2022}, which is the necessary condition in the model. How compressional heating \citep{Townsley2004ApJ...600..390T} might disrupt the magnetic field emergence is another aspect under consideration, given the estimated temperatures for the known WD pulsars of $\approx 11,500$\,K \citep{Garnavich_ARSco_UV:2021ApJ...908..195G, Pelisoli_J1912_UV:2024MNRAS.527.3826P}. Moreover, the recent identification of M-dwarf – WD binaries in some long-period radio transients (LPTs) suggests a possible evolutionary link to WD pulsars \citep[][]{Hurley-Walker_LPT:2024ApJ...976L..21H,deRuiter_LPT:2025NatAs.tmp...78D,Rodriguez_LPT:2025A&A...695L...8R}.

Despite their undisputed importance, a larger sample of WD pulsars must be identified in order to draw an evolutionary scenario. Here we present the discovery of the third system of this class, \srclong\ (hereafter \src). We make use time-resolved spectroscopy and high-cadence photometry to determine the fundamental properties of this new system and compile the observational properties of all three known WD pulsars, providing a benchmark for the classification of future candidates.

\begin{figure*}
      \centering
      \includegraphics[width=0.98\textwidth]{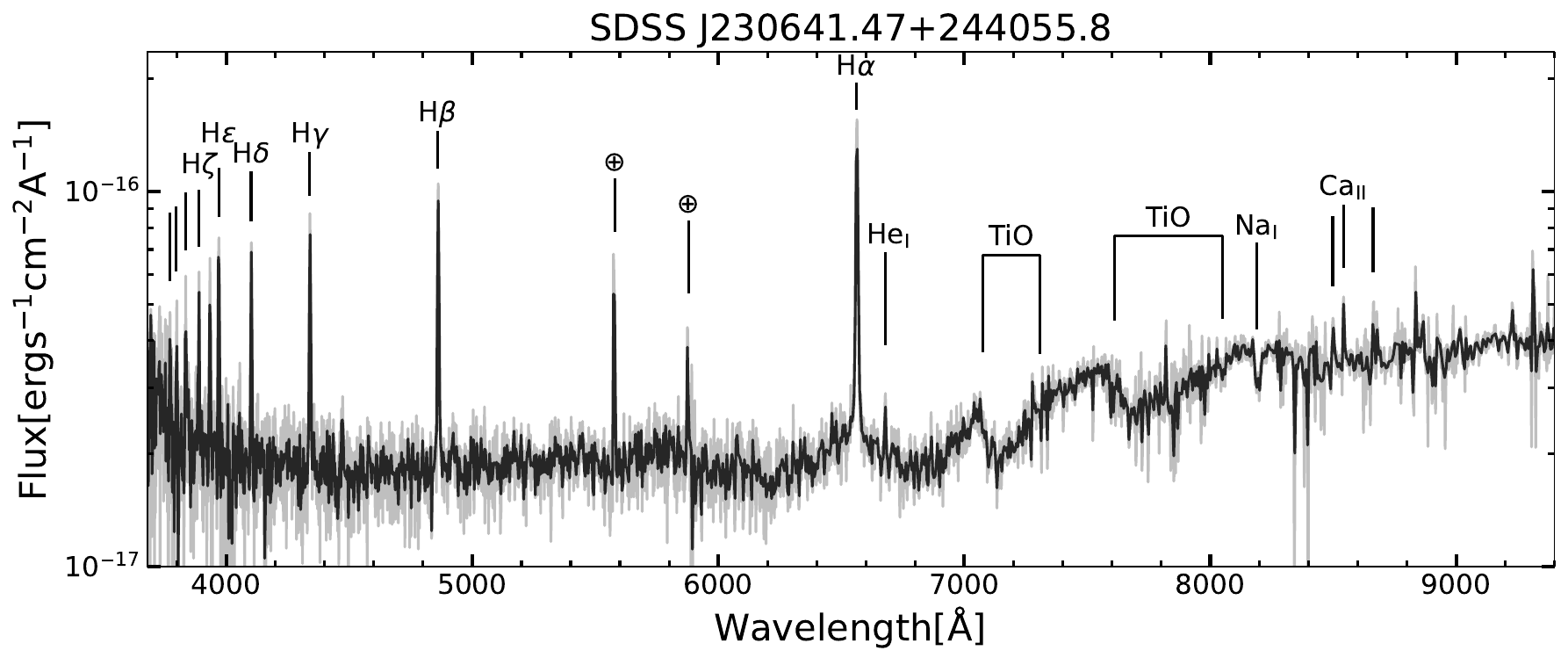}
    \caption{{SDSS} optical spectrum of \src. The unbinned and binned (by a factor of three) spectrum is shown in a light grey solid and black solid lines, respectively. The Balmer series is identified, superimposed on a relatively featureless continuum. At wavelengths longer than 6000\,\AA, the spectrum exhibits TiO molecular bands typical of an M-dwarf atmosphere; the lines in the red side of the spectrum that are studied in the present research are also labelled (i.e. H$_\mathrm{\rm \alpha}$, He\,$\lambda 6678$\,\AA, the Ca {\sc ii} triplet ($\lambda\lambda 8498,\, 8542{\, \rm \&\,} 8662$\,\AA) and the 8190\,\AA~Na\,{\sc i} doublet. The SDSS identifiers of this spectrum are: MJD=57332, plate=7705, fiber=321.}
    \label{fig: full_spectrum}
\end{figure*}

\section{Data Sets} \label{sec: observations}

\subsection{Initial discovery data}
\src\ was classified as an EW-type contact binary with an orbital period of 6.988\,h by \citet{Chen2020ApJS..249...18C} from an analysis of the  Zwicky Transient Facility  \citep[ZTF; ][]{ZTF_2019PASP..131a8002B,ZTF2_2019PASP..131f8003B,ZTF3_2019PASP..131g8001G,ZTF4_2020PASP..132c8001D} Data Release (DR)~2 photometry. Later, \citet{Inight_SDSS_CVs:2023MNRAS.524.4867I} identified \src\ as a CV candidate on the base of its Sloan Digital Sky Survey  \citep[SDSS;][]{SDSS_2017AJ....154...28B} spectrum, noting the dominant contribution of the companion star.
As we show in Fig.\,\ref{fig: full_spectrum}, \src's spectrum gathered by SDSS exhibits strong narrow emission lines in the Balmer series overlaid on top of a relatively feature-less continuum in the blue-end. The red-end of the spectrum, on the other hand, is dominated by molecular bands characteristic of cold M-dwarf companions. This is one of the key characteristics observed in WD pulsars (see Sec.\,\ref{sec: wdp discussion}).

\subsection{Time-resolved photometry}

In data release 20 (DR20), The Zwicky Transient Facility \citep[ZTF; ][]{ZTF_2019PASP..131a8002B,ZTF2_2019PASP..131f8003B,ZTF3_2019PASP..131g8001G,ZTF4_2020PASP..132c8001D} gathered long term photometry of \src\ ranging from 2018/05/08 to 2023/10/25, yielding a total of 1473 individual exposures (436, 857 and 180 for $g$, $r$ and $i$ respectively). ZTF is a wide-field survey covering $47\,{\rm deg^2}$ of the sky. It is mounted on the {Samuel Oschin 48'' Schmid telescope at Palomar Observatory}. The nominal exposure time is 30\,s, reaching a $5\sigma$ detection limit of 20.5 mag. We used this data to constrain the orbital period of the system (see Sec.\,\ref{sec: Porb}). The folded light curve for the $r$ band is presented in the top panel of Fig.\,\ref{fig: folded lc and rv}. 

We observed \src\ with ULTRASPEC \citep{ULTRASPEC}, a high-speed photometer mounted at the 2.4-m Thai National Telescope (TNT). ULTRASPEC is equipped with a $1024\times1024$ frame-transfer, electron-multitplying CCD (EMCCD) with a maximum field of view of $7.7\times7.7\,{\rm arcmin^2}$. We executed four visits to \src\ during the nights of the 13, 14, 15 December 2022 and 29 November 2023 with exposure times of $8$ s, $15$ s, $8$ s and $15$ s respectively, with only 15 ms
dead time between each exposure. All observations used a wide KG5 filter \citep[effectively $u + g + r$;][]{Hardy_USPEC_GK5band_2017MNRAS.465.4968H}. A full log of the ULTRASPEC observations is listed in Table\,\ref{tab: uspec log}. The data were reduced using the \texttt{HiPERCAM} pipeline\footnote{https://github.com/HiPERCAM/hipercam} \citep{Hipercam:2021MNRAS.507..350D}.  Prior to analysis, all the bad intervals due to poor weather conditions were removed and all the timestamps were centred to mid exposure and converted to the barycentric reference frame using \texttt{jpl} ephemerides by applying the methods implemented in \texttt{astropy.time} \citep{Astropy2018AJ....156..123A}.

\begin{table}
	\centering
	\caption{ULTRASPEC observing log.}
	\label{tab: uspec log}
	\begin{tabular}{lcccr} 
		\hline
		Epoch & MJD$_{\rm start}$ & filter & exposure time & obs. length\\
		\hline
        \ & day & \ & second & hours
        \\
        \hline
        $1$& $ 59926.499 $& $KG5$ & $7.8$ & $3.52$ \\
        $2$& $ 59927.511 $& $KG5$ & $15.8$ & $2.90$ \\
        $3$& $ 59928.487 $& $KG5$ & $7.8$ & $3.62$ \\
        $4$& $ 60277.487 $& $KG5$ & $15.8$ & $2.56$ \\
        \hline
	\end{tabular}
\end{table}

\subsection{Time-resolved spectroscopy}

We obtained long-slit spectroscopy of \src\ with the {Gemini Multi-Object Spectrograph} \citep[GMOS;][]{GMOS-N2004PASP..116..425H} mounted on the 8-m Gemini North telescope (program ID: GN-2023B-Q-228). 
The GMOS instrument is equipped with two Hamatsu CCDs ($2\times 2048 \times 4176$ pixels). Time-resolved spectroscopy was obtained in two visits during two different nights (on the 28th \& 30th July 2023), yielding a total of 20 individual exposures. 
Each night, two consecutive observations (OBs) were performed, comprising one arc and four science exposures, designed to minimize the impact of flexures on the wavelength calibration.
The exposure time ($t_\mathrm{exp} = 553\,{\rm s}$) was selected to be a multiple of the period of the high frequency signal to dilute any impact in the line profiles. 

The observations were carried out with the R831 grating centred at {750\,{\rm nm}}, $2\times2$ binning, and $1\,{\rm arcsec}$ slit width, providing coverage of H$_{\rm \alpha}$, He\,$\lambda 6678$, the 8190\,\AA~Na\,{\sc i} doublet, and the Ca {\sc ii} triplet ($\lambda\lambda 8498,\, 8542{\, \rm \&\,} 8662$\,\AA), with a resolving power $R\simeq 4400$, equivalent to a resolution of $\delta v\simeq68\,{\rm km\,s^{-1}}$. The data was reduced using standard procedures using the latest version of the \texttt{DRAGONS} pipeline\footnote{https://dragons.readthedocs.io/en/stable/}. Barycentric correction was applied to the timestamps using the same method as above. The spectroscopic data were also corrected to the barycentric rest frame.

\section{Analysis and results}\label{sec:analysis}

\begin{figure*}
      \centering
      \includegraphics[width=0.88\textwidth]{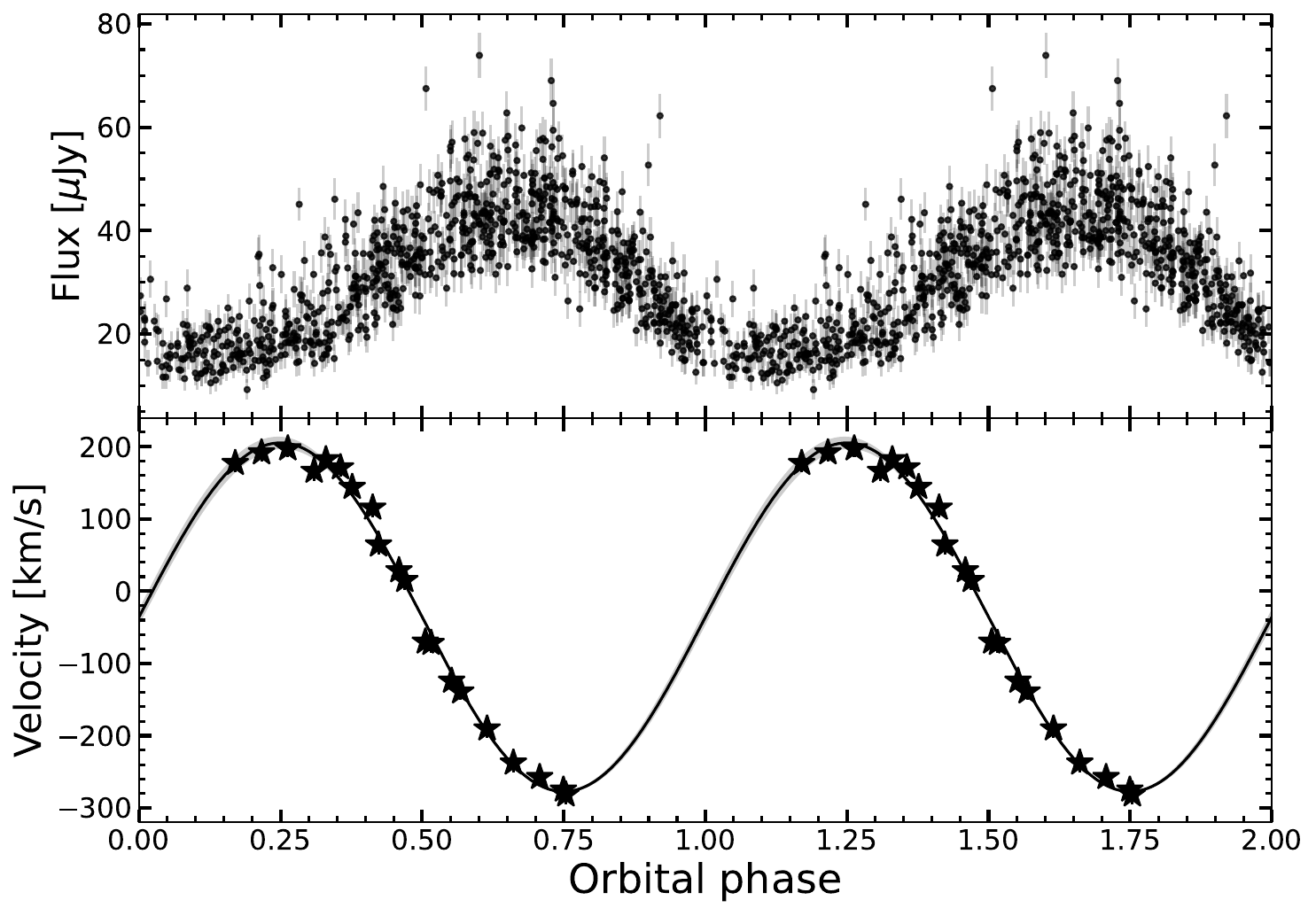}
    \caption{Optical flux and radial velocity curve of \src\, as a function of the orbital phase. \textbf{ Top:} ZTF optical light curve in the $r$ band. The orbital modulation exhibits a large scatter at orbital phases when the trailing face of the companion star faces Earth similar to what is observed in \arsco\ and \wdp\ \citep[e.g. ][]{Marsh2016Natur.537..374M, Pelisoli2023NatAs...7..931P}. \textbf{ Bottom:} Sinusoidal modulation of the Na {\sc i} absorption lines tracing the orbital motion of the companion M star with a period of $3.49$ h.}
    \label{fig: folded lc and rv}
\end{figure*}

\subsection{Photometric monitoring}
\subsubsection{Orbital period} \label{sec: Porb}

To characterise the long term variability of \src\ we combined all the data available from ZTF to compute the multi-band generalized Lomb-Scargle periodogram \citep[][]{Lomb1976,Scargle1982,VanderPlas_multi-band_LS:2015ApJ...812...18V,vanderplas2018} as implemented in \texttt{astropy} version 6.1.2. We obtain a clean detection of a signal $ P \simeq 3.5\,{\rm h}$ in the resulting power spectrum associated with the orbital period of the system. 
The data also showed a hint of a short period which further motivated the follow up. 
We determined the central frequency of the main signal and its uncertainty using a bootstrap analysis \citep[e.g. ][]{Stats_ML_DataMining2014sdmm.book.....I}, resulting in  

$$\rm P_\mathrm{orb} = 3.4939558(3)\,h.$$

Interestingly, this value is half of the period that triggered the classification as a contact binary by \citet{Chen2020ApJS..249...18C}. The long-term photometric (folded) light curves are presented in the top panel of Figures \ref{fig: folded lc and rv} and \ref{fig: folded lc} for the $r$-band and $gri$-bands respectively. The orbital modulation from the companion star dominates the $g$ and $r$ bands as shown in these figures, as a result of an increase in the effective temperature of the companion star facing the WD, which is being heated by radiation, interaction of the magnetic fields from the two binary components, or both. (Sec.\ref{sec: spectroscopy}). 
Superimposed on the orbital modulation, we can see a variable scatter with larger amplitude at specific orbital phases. These short term variations are intrinsic to the source, resembling the behaviour seen in \arsco\ and \wdp\ \citep[e.g. ][]{Marsh2016Natur.537..374M, Pelisoli2023NatAs...7..931P}.

\subsubsection{Spin period} \label{sec: Pspin}

\begin{table}
	\centering
	\caption{List of all the coherent signals extracted from ULTRASPEC data. $\omega$ is the strongest signal in the power spectrum  which we associate with the spin period. Combinations of the beat between the orbital frequency $\Omega$ with $\omega$ are also identified.}
	\label{tab:spin}
	\begin{tabular}{lcr} 
		\hline
		Frequency (mHz) & Period (s) & Origin\\
		\hline
        $10.836(3)$ & 92.28(3) & $\omega$\\
        $21.685(4)$ &  46.11(1) & $2\,\omega$\\
        $21.853(5)$ & 45.76(1) & $2\,\omega + 2\,\Omega$\\
        $32.62(2)$ & 30.65(2) & $3\,\omega + \Omega$\\
        \hline
	\end{tabular}
\end{table}

\begin{figure*}
      \centering
      \includegraphics[width=0.88\textwidth]{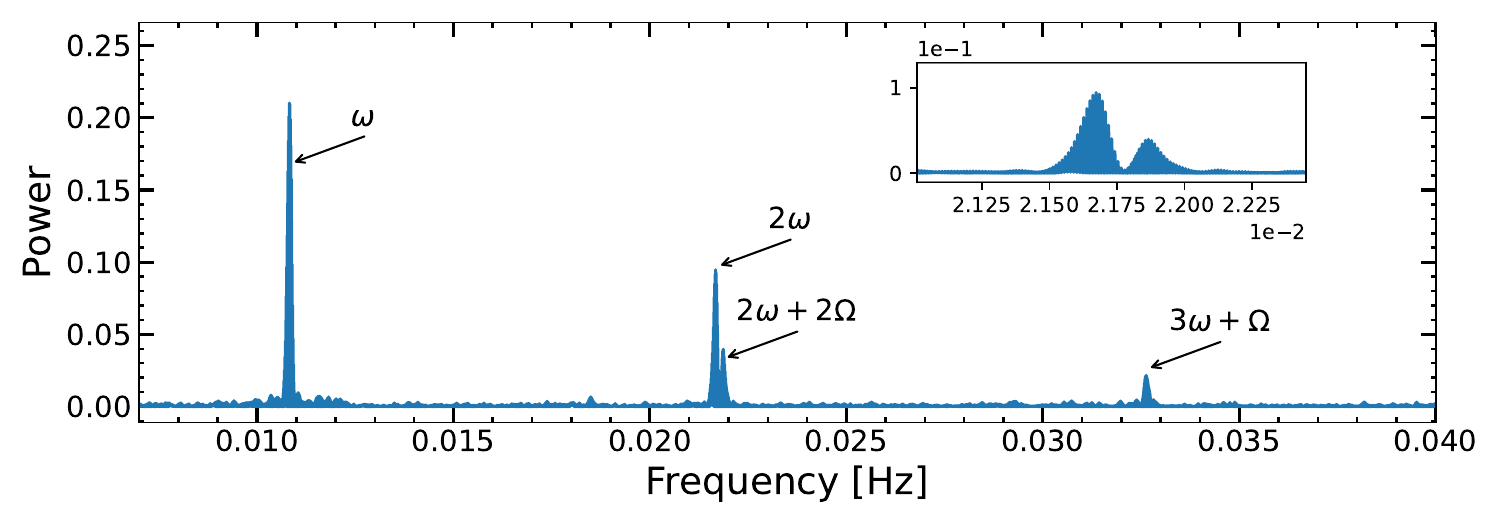}
    \caption{Power spectrum of the high cadence data obtained with ULTRASPEC (after removing the orbital modulation). The strongest signal is at $\rm P_\omega \simeq 92$ seconds, the first harmonic of this signal is clearly identified, as well as the beat between the orbital period ($\Omega$) and $\rm P_\omega$. The inset shows a zoom in around $\rm 2\times \omega$ illustrating the excess of power on the beat period ($\rm 2 \omega + 2\Omega$).}
    \label{fig: uspec power}
\end{figure*}

\begin{figure}
      \centering
      \includegraphics[width=0.49\textwidth]{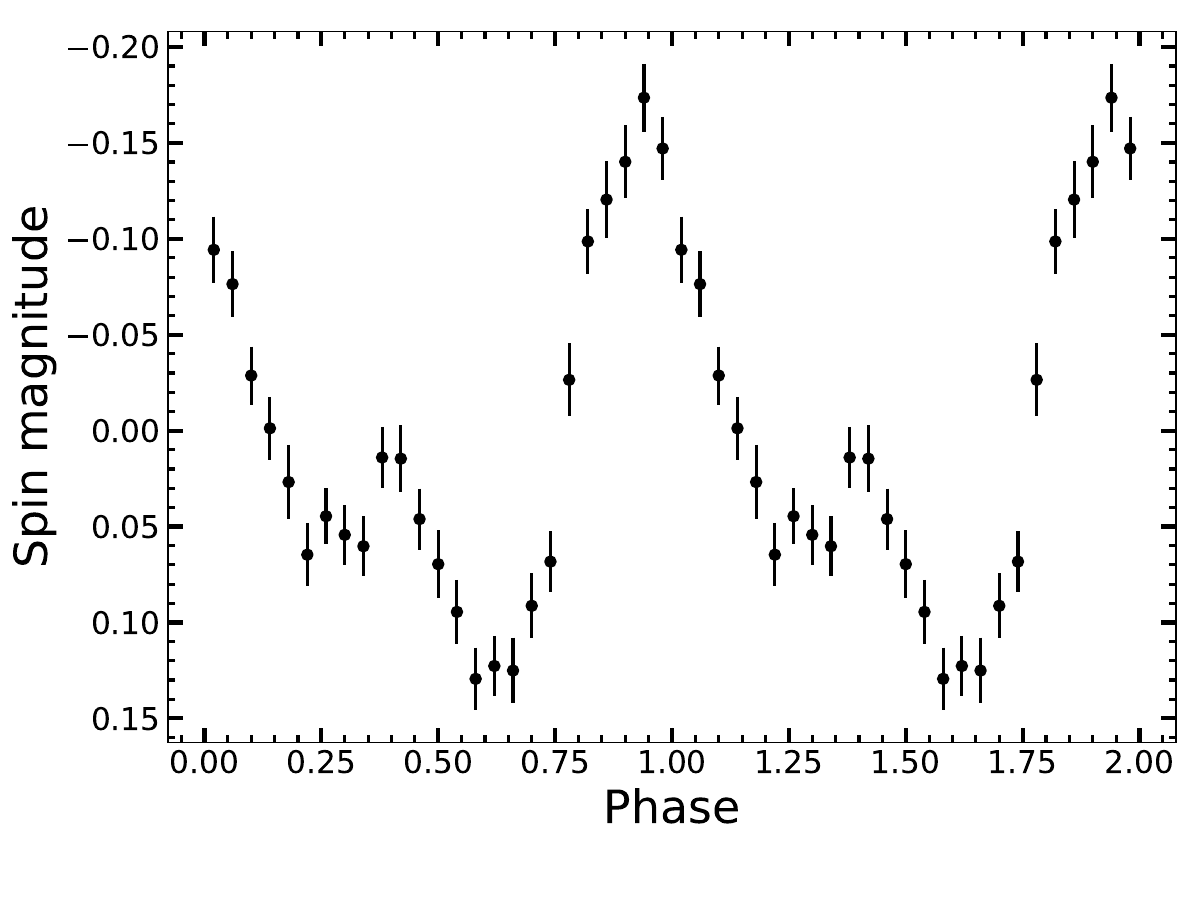}
    \caption{ULTRASPEC light curve of \src\ folded on the $\rm P_\omega \simeq 92$\,s period. The orbital modulation was subtracted from the data before folding. The resulting light curve exhibits two distinct pulses per cycle, potentially owing to the change in the visibility of the near- and far-side poles of the WD. The spin phase ephemeris is arbitrary and two cycles are shown for clarity.}
    \label{fig: folded spin}
\end{figure}

We have used the fast photometry gathered with ULTRASPEC to determine the short-term properties of \src. To remove the orbital modulation before proceeding with any analysis, we de-trended the observed light curves using a sinusoidal function with the orbital frequency derived above. We then computed the Lomb-Scargle (LS) power for the combined time series and group by exposure times, fast ($t_\mathrm{exp}=7.8$ s) and slow ($t_\mathrm{exp}= 15.8$ s). Finally, we determined the uncertainties for the main signals in each of these. The nights with longer exposures not only have slightly less accumulated exposure time but also less favourable weather conditions, yielding less precise results when these data are incorporated to the analysis, i.e. we recover the same frequencies with higher uncertainty. Thus, we only retain the higher cadence in our analysis. The resulting power spectrum is shown in Fig.\,\ref{fig: uspec power}, where we can see a dominant signal of $\rm P_\omega \simeq 92\,s$, which we interpret as the spin of the WD producing the pulsed emission as the viewing angle of the WD changes and/or its magnetic field interacts with the companion star in each rotation, provided that is the only detected signal at low frequencies. We can also see some of its respective harmonics, as well as some of the beat periods between this signal ($\omega$) and the orbital period. To determine the centroid and $1\sigma$ errors we followed the same bootstrapping method as in Sec.\,\ref{sec: Porb}. The list of all the coherent signals detected in the high-speed photometry and their proposed origin are summarized in Table \ref{tab:spin}.

In Fig.\,\ref{fig: folded spin} we show the light curve resulting from subtracting the orbital modulation folded into $P_\omega$ and rebinned. This figure reveals a double pulsed emission within one phase with a primary pulse amplitude of $\sim 6$ per cent. However, the intrinsic shape of the signal may have been distorted given the combination of low S/N with possible variations in amplitude and shape of pulsed emission with the orbital phase and on longer timescales \citep[e.g. ][]{Pelisoli_ARSco_long-term:2022MNRAS.516.5052P}. 
The modulation of the pulse amplitude induced by $\rm P_\omega$ is seen in the upper panel of Fig.\,\ref{fig: folded lc and rv}. 
To quantify this change, we separate the light curve (folded with $\rm P_\mathrm{orb}$) into 10 segments, thus, covering $\Delta\phi_\mathrm{orb} = 0.1$ per segment. We perform a sinusoidal fit to each of these segments using the Simplex algorithm \citep[][]{simplex_NelderMead:1965,simplex_wright:1996} as implemented in \texttt{scipy} version 1.13.0. 
Using all the high-cadence data from ULTRASPEC, we obtain a typical peak-to-peak variability amplitude of $\lesssim 10$ per cent reaching a minimum amplitude of $\sim 7$ per cent variability during certain orbital phases. The signal to noise ratio of the fast photometry prevents us from quantifying these changes precisely, but qualitatively we can say that the dispersion decreases towards $\phi_\mathrm{orb}= 1$, i.e. when the trailing face of the companion star is on the far side, which is consistent with the spectroscopic modelling below.

\subsection{Spectroscopic analysis} \label{sec: spectroscopy}

\begin{figure*}
      \centering
      \includegraphics[width=0.24\textwidth]{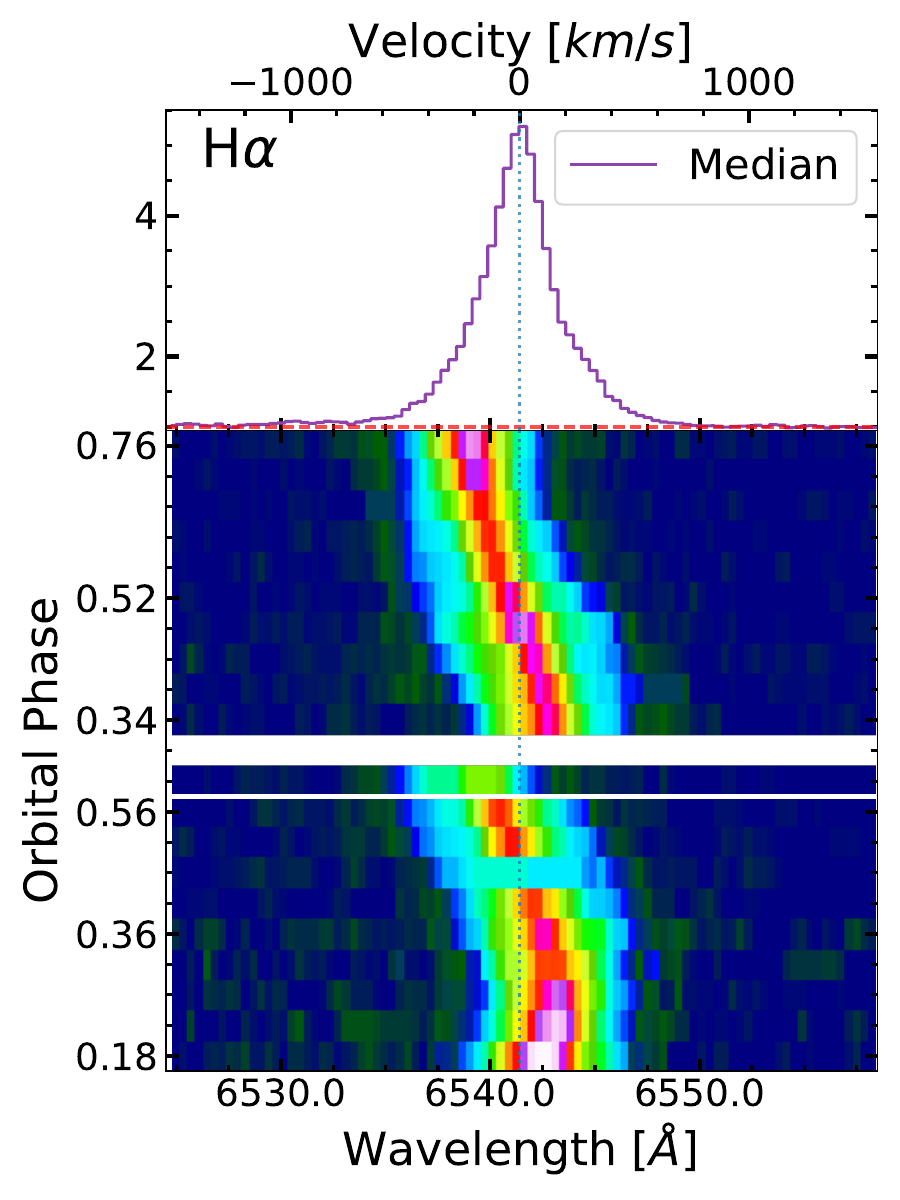}
      \includegraphics[width=0.21\textwidth]{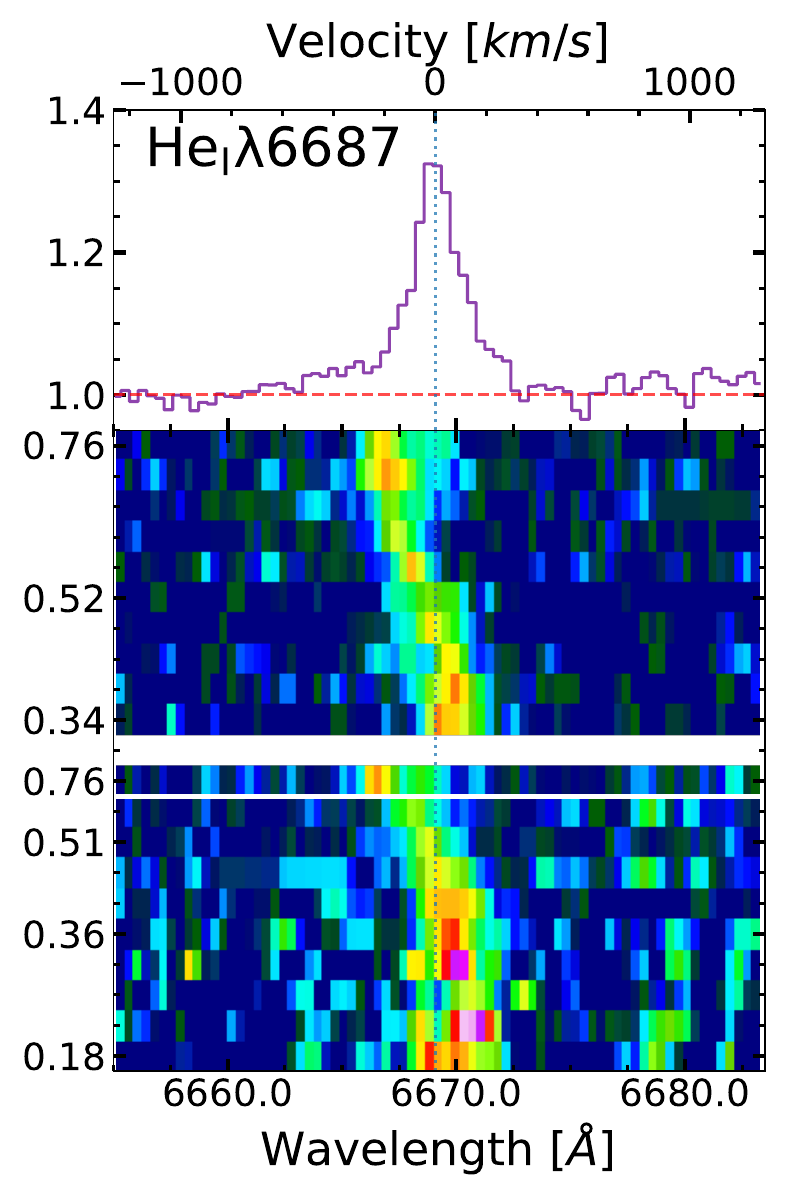}
      \includegraphics[width=0.32\textwidth]{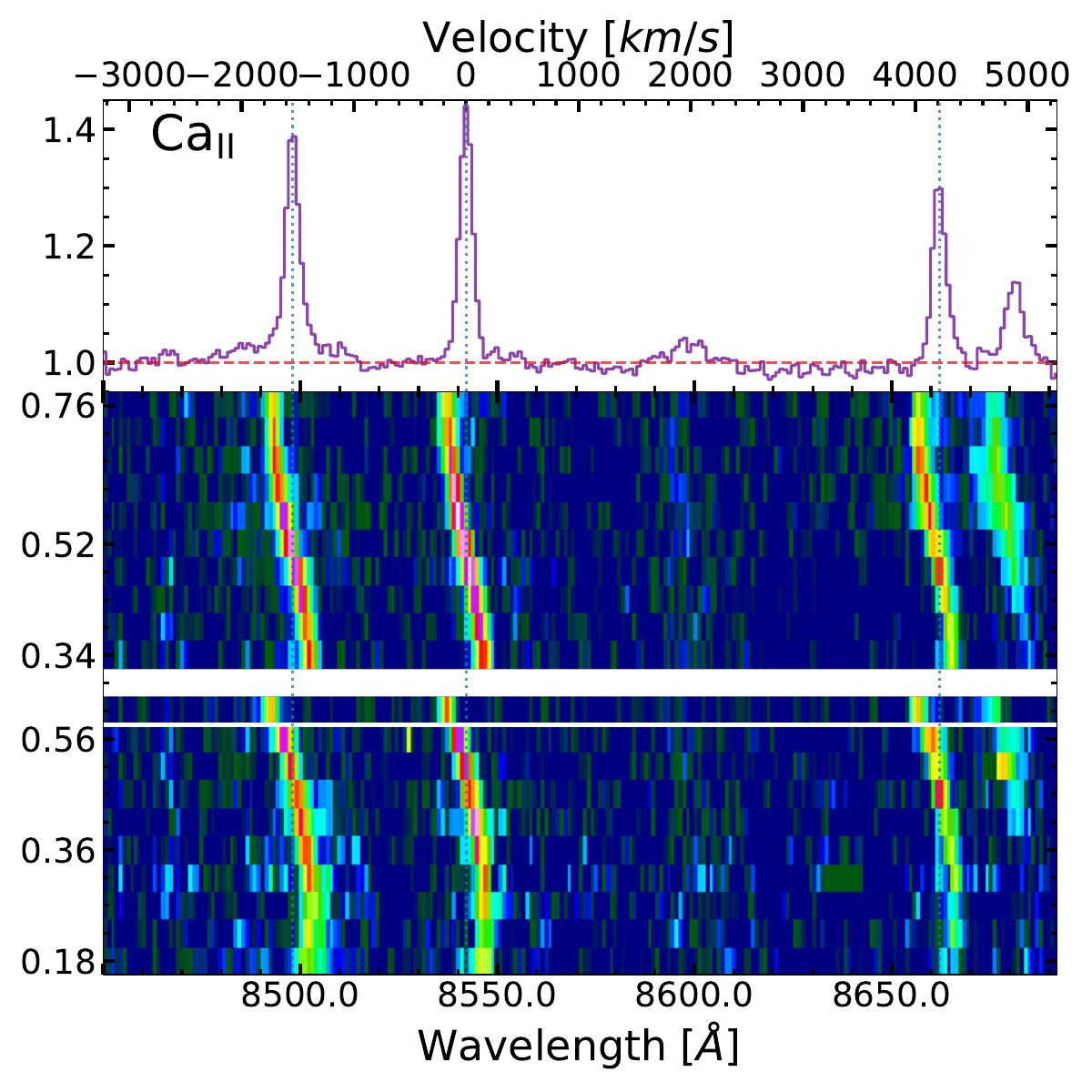}
        \includegraphics[width=0.21\textwidth]{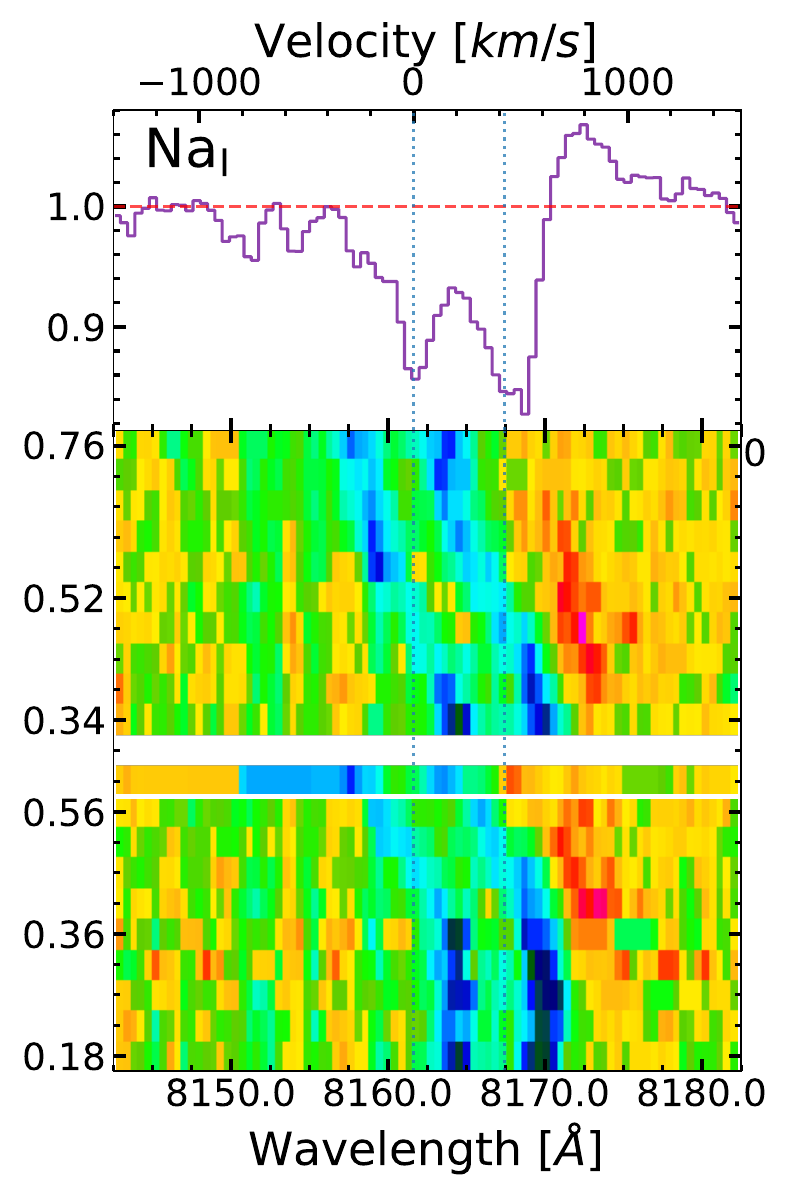}
    \caption{Trailed spectra of the spectral lines analyzed in the present research. From left to right: H$_{\rm \alpha}$, He\,{\sc i}$\lambda 6678$, Ca {\sc ii} triplet ($\lambda\lambda 8498,\, 8542{\, \rm \&\,} 8662$\,\AA), and the 8190\,\AA~Na\,{\sc i} doublet. The top panels show the rest-position median spectrum for each line. The bottom panels show the time-resolved trailed spectra for the two sets of observations separated by a horizontal white space after $\phi_\mathrm{orb}=0.76$, the horizontal line in the bottom set indicates a gap in the sequence, the specific orbital phase for the last observation of that set is 0.76. 
    The typical phase difference between consecutive exposures is $\rm \Delta \phi = 0.046$. In these panels the intensity is colour coded as dark blue corresponding to the minimum and light magenta to the maximum. Note that some of the individual exposure have masked pixels due to the impact of cosmic rays in the CCD during the exposure.} 
    \label{fig: trailed}
\end{figure*}

The time-resolved spectroscopy was obtained during two consecutive nights with a small gap during the first visit and partial overlapping of the orbital phase. 
In Fig.\,\ref{fig: trailed} we show the trailed spectra of all the lines of interest, namely  H$_{\rm \alpha}$, He\,$\lambda 6678$\,\AA, the Ca {\sc ii} triplet ($\lambda\lambda 8498,\, 8542{\, \rm \&\,} 8662$\,\AA) and the 8190\,\AA~Na\,{\sc i} doublet. 
In this section we present an in-depth analysis of the spectroscopic properties of \src.
For our analysis we have used the spectroscopic data binned to two pixels per resolution element determined by the resolving power of our instrument setup (i.e. $\rm R\simeq 4400$). 

\subsubsection{Line Modelling \& Orbital Ephemerides}

From Fig.\,\ref{fig: trailed} it is clear that \src\ is a spectroscopic binary as it exhibits S-waves in the Hydrogen and Helium lines, as well as the Calcium triplet in emission. These are typical signatures observed from the inner face of companion stars owing to the temperature inversion in their surface due to the heating caused by the compact object or magnetic reconnection \citep[e.g. ][]{Haefner_NNSer:2004A&A...428..181H, Garnavich_ARSco_time_resolved_sp:2019ApJ...872...67G}.

To obtain the radial velocity curves of the emission lines, we modelled the centroid of the lines.
We performed all the fits to the continuum-subtracted spectra using models implemented in \texttt{astropy} and carrying out Markov chain Monte Carlo (MCMC) with a Goodman \& Weare's Affinite Invariant Ensemble sampler as implemented in {\sc emcee}\footnote{https://github.com/dfm/emcee} \citep{Astropy2018AJ....156..123A,emcee2013PASP..125..306F}. For each line, we first modelled the underlying continuum using a second order polynomial fitted to the neighbouring regions where contribution from the lines is negligible. The lines were modelled using Gaussian profiles with an added constant, to avoid bias owing to over/under estimation of the continuum at the line centre. Simple non-informative uniform priors were adopted for all fit parameters. To produce the MCMC chains we used 100 walkers per free parameter and ran them until the chains were sufficiently converged, i.e. until the chain number of iterations was at least $100\times$ the auto correlation time ($\tau$) and the latter did not improve more than 1 per cent after 500 iterations. The first $2 \times \tau$ iterations were discarded and the chains were thinned by segments of $\tau/2$ to minimise the effect of auto-correlation from the sampler. For all the fits, the resulting posterior distributions exhibit roughly normal distribution for all the parameters. 
The best fitted values are drawn from the posterior probability distributions of each parameter with the uncertainties drawn from the 16 per cent \& 84 per cent quantiles of the sampled posterior.

In the case of H$\alpha$ and He\,{\sc i}, a single Gaussian component does not provide a good fit to the line profile in most cases. 
This is due to the influence of a broad base which is seen in the median spectra of Fig.\,\ref{fig: trailed}. 
We attempted to fit the profiles with two Gaussian components, however, this resulted in poorly constrained and unphysical values, even when forcing the centre of one of the narrower components to the centroid of the emission core fitted with a second order polynomial. 
Therefore, we limited our analysis of H$\alpha$ and He\,{\sc i} to determining only the centroid of the narrow emission by fitting a second order polynomial to the maximum of the line. In this case, we determined the error on the lines by bootstrapping each wavelength bin assuming Gaussian errors determined from the adjacent continuum.  
On the other hand, the Ca triplet emission lines are not resolved with our spectral resolution. The triplet was fitted with its full width at half maximum (FWHM) tied to a single value in each epoch. However, the radial velocity was required to have some degree of freedom to accommodate the fit in some of the individual exposures due to the systematics introduced during the wavelength calibration. 
Finally, we modelled the Na\,{\sc i} absorption doublet at $\lambda\lambda=8150, 8224$\,\AA, shown in the right panel of Fig.\,\ref{fig: trailed}. In this case, we used two Gaussians with FWHM tied to the same value and the distance between them was fixed to their laboratory separation.

In order to determine the time 0 ephemerides ($T_0$) of \src, defined as the inferior conjunction of the companion, we perform a sinusoidal fit to the centroid of each ionisation species with an expression of the form:
\begin{equation} \label{eq: RVs}
    v_i(t) = \gamma_i + K_i \sin{(2\pi(t - T_0)/P_{\rm orb})},
\end{equation}
where $v_i(t)$ and $K_i$ are the observed velocity and its amplitude for the $i$-th ionisation species.
$\gamma_i$ is the systemic velocity of such species, and $\rm P_\mathrm{orb}$ is the orbital period determined from the long-term photometric light curves (Sec. \ref{sec: Porb}). 

\begin{figure}
      \centering
      \includegraphics[width=0.48\textwidth]{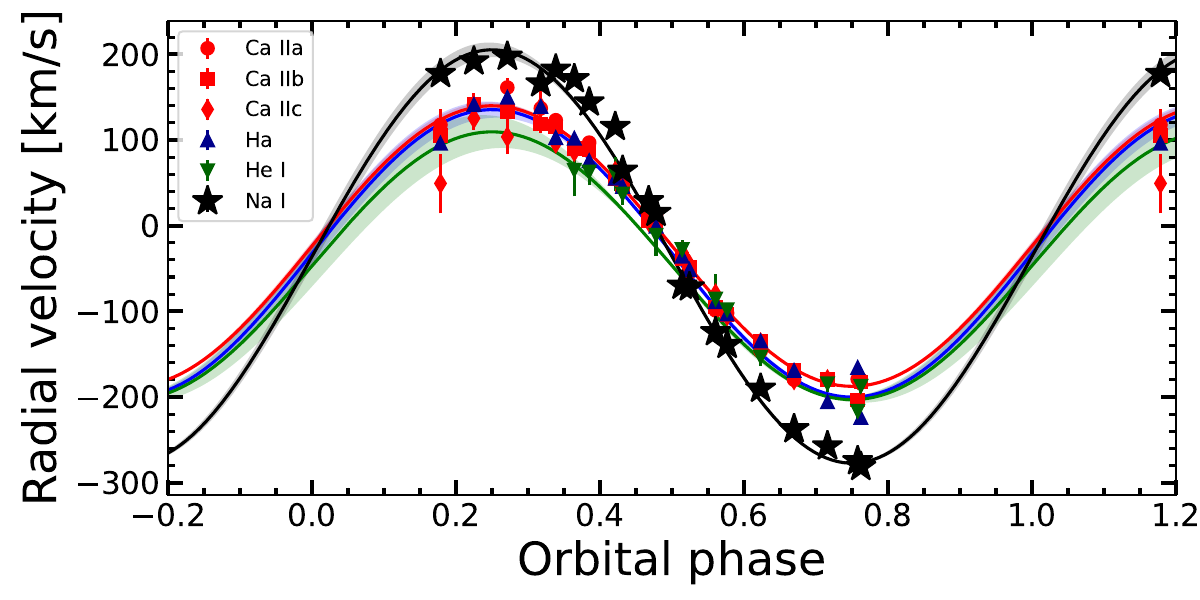}
    \caption{Radial velocities folded with the orbital phases. Each symbol corresponds to an absorption or emission line measured from the time resolved {GMOS} spectroscopy. The solid line is the best sinusoidal fit to each ionisation species and the shaded regions represent the $\pm 1\sigma$ error on the fit. Error bars in Na\,\textsc{i} are smaller than symbols.} 
    \label{fig: folded rvs}
\end{figure}

The radial velocity measurements for each spectroscopic line along with the the best fits to equation \ref{eq: RVs} are shown in Fig.\,\ref{fig: folded rvs}, and in the bottom panel of Fig. \ref{fig: folded lc and rv} for Na\,{\sc i}. The fit was performed using the MCMC technique described above. 
In this case, the uncertainty in the orbital period was modelled as a Gaussian distribution, with both the mean and standard deviation treated as a hyperparameter in the MCMC chains.  
A total systematic error ($\sigma_{\rm sys}$) was also introduced in the likelihood function as a free parameter to capture errors introduced by small offsets in the wavelength solution during the data reduction, so the uncertainty for the $n$-th radial velocity measurement can be written as $\sigma_\mathrm{T,n}^2 = \sigma_n^2 + \sigma_{\rm sys}^2$. We have estimated the $T_0$ ephemerides for each of the lines and the combination of them resulting in consistent values for all of these. Given the uncertainties introduced by the low S/N on He\,{\sc i}, and the fact that in some of the epochs H$\alpha$ are affected by cosmic rays, here we report the ephemerides with the least uncertainty obtained from the calcium triplet:

$$\rm T_0^{Ca\,_\mathrm{\textsc{ii}}}(MJD_{TDB}) = 60154.2911(4) + 0.14558149 (1) E .$$

\subsubsection{The Companion Star \& Mass Function} \label{sec: mass function and companion}

As shown in Fig.\,\ref{fig: full_spectrum}, the red spectrum of \src\ unambiguously identifies the companion as an M-dwarf, exhibiting molecular bands characteristic of these late-type stars. Our time-resolved spectroscopic observations are designed to cover this spectral region (spanning $\lambda\simeq 6320 - 8700$\AA), which contains key information about the stellar component in the binary system. 
The sodium doublet is particularly important as it approximately traces the centre of mass of the companion star. It is worth noting that the centre of light for the sodium doublet can be offset by the difference in temperature difference between the star’s hemispheres  \citep[c.f. ][]{Friend_NaI_shift:1990MNRAS.246..654F}. Our data is not exempt from this effect, as evidenced by the weakening of the absorption feature around phase 0.5 in Fig.\,\ref{fig: trailed}, which makes our constrains on $K_2$ effectively an upper limit. However, since the best-match template does not change with the orbital phase (see below), the impact of this effect in our case must be small and is likely dominated by other sources of systematic error. With this in mind, we use the semi-amplitude of the Na\,\textsc{i} line to constrain the binary mass function of the system ($f(M_1)$, following the same approach as \citealt{Marsh2016Natur.537..374M} and \citealt{Pelisoli2023NatAs...7..931P}). The mass function can be written as
\begin{equation} \label{eq: f(m)}
    f(M_1) = \frac{M_1^3\sin^3{i}}{(M_1+M_2)^2} = \frac{K_2^3P_\mathrm{\rm orb}}{2 \pi G}
\end{equation}
where $P_\mathrm{orb}$ is the orbital period, $K$ is the radial velocity semi-amplitude tracing the centre of mass of the donor, $G$ is the gravitational constant, $i$ is the orbital inclination, $M_1$ and $M_2$ are the mass of the primary and the secondary respectively. 
Combining the values obtained above we obtain $f(M_1)= 0.21\pm0.01\,{\rm M_\odot}$, where we used $K=K_\mathrm{Na\,\textsc{i}}$.

The Na\,{\sc i} line is also very sensitive to the effective temperature of the stellar atmosphere and hence to its spectral type. The presence of this line itself implies an M-dwarf companion. 
To infer the spectral type, we have compared our {GMOS} data against a set of well-characterised M-dwarf spectral templates presented in \citet{Parsons2018:MNRAS.481.1083P}. 
The best match was determined by computing the $\chi^2$ of the red-end spectrum dominated by the M-dwarf, using only regions free from either telluric or emission lines\footnote{i.e. excluding $\lambda\lambda = 6310- 6600, 6664- 6690, 7088- 7118, 7348- 7410, 7587- 7742, 7886- 7918, 8460- 8700$\AA.}. 
Prior to the comparison we de-reddened all the individual spectra using the \citet{Fitzpatrick:1999PASP..111...63F} extinction law as implemented in  \texttt{dust\_extinction}\footnote{https://dust-extinction.readthedocs.io} v1.5.
The extinction was determined using the 3D maps from \citet{Bayestar:2019ApJ...887...93G}, these maps reveal a constant extinction of $E(g-r) = 0.23\pm 0.02$ in the line of sight throughout the region populated with stars (i.e. from $\simeq0.5\,\rm kpc$ to $\simeq6.5\,\rm kpc$). $
E(B-V)$ was derived using the values tabulated in Table 6 of  \citet{SFD_extinction:2011ApJ...737..103S} and $R_V = A_V/E(B-V) = 3.1$ \citep[e.g. ][]{Savage&Mathis:1979ARA&A..17...73S}.
We then centred the observed spectra in the donor rest frame using the radial velocities from Sec.\,\ref{sec: spectroscopy}. 
To account for other spectral components that may influence the shape of the underlying continuum, 
we fitted each template to the observed spectra with a veiling component emulating the non-thermal emission which is represented by a power law  (using least squares fits as implemented in \texttt{astropy}). 
The data is best described by a M$2.5$ spectral type, for the median spectrum as well as the individual exposures. This value is earlier than the M$4.5$ spectral type observed for other WD pulsars and may be underestimated due to the lack of spectra at 0 orbital phase. 
To test this further, we have repeated the same experiment using the SDSS spectrum shown in Fig.\,\ref{fig: full_spectrum}. Using this data, the best match is obtained for the M$4$ with similar values for M$3$ and M$5$ the spectral template matching with the. These values are a more natural match for \src's orbital period and are consistent with fact that all three WD pulsars are clustered in the same region of the colour magnitude diagram (see Sec.\,\ref{sec: wdp discussion}). Therefore, we conclude that the most likely spectral type of \src\ is M$4.0\pm0.5$.
From the individual fits of the GMOS data, it is also clear how the veiling factor changes in each orbital phase. Even though we do not cover all the orbital phases, in these fits the veiling factor decreases rapidly from $\phi_\mathrm{orb} \simeq0.17$ finding a minimum around $\phi_\mathrm{orb} \simeq0.6$, consistent with the phase dependent scatter from the light curve estimated in Sec.\,\ref{sec: Pspin}.

By subtracting the non-stellar continuum from the best fitted template, we can obtain a good estimation of the stellar flux that can be used to further characterize the stellar component. 
We used the grid of stellar M-dwarf models from \citet{NextGen_Allard:1997ARA&A..35..137A} to compute the maximum likelihood function of the effective temperature ($\rm T_\mathrm{eff,2}$), and the normalization factor $\log_\mathrm{10}{(R_2/d)^2}$, where $R_2$ and $d$ are the radius of the stellar companion and the distance to the system respectively. 
We obtain the most likely values and their equivalent $1\sigma$ uncertainties to be $\rm T_\mathrm{eff} = 3600\pm100$\,K and $\log_\mathrm{10}{(R_2/d)^2}= -21.73\pm0.06$. We note that a lower temperature would be more typical for this class of system and spectral type, we suspect that it may have been slightly overestimated for the same reasons discussed above. Repeating the same logic as with the spectral type classification, we find that an effective temperature of $3500$\,K have the most likelihood, however $\rm T_\mathrm{eff} = 3300\pm100$\,K lead to similar $\chi^2$ while being consistent with the value quoted above and providing a more natural match to the spectral class.
All the values are summarised in Table \ref{tab: params}.

\begin{table} 
	\centering
	\caption{System parameters of \src. \textsuperscript{\textdagger} indicates estimates derived assuming that the average luminosity in the Gaia $G$-band is similar to the one from \arsco\ and adopting the semi-empirical mass-radius relation from \citet{Brown2022MNRAS.513.3050B}, see Sec.\,\ref{sec: distance} for details. 
    }
	\label{tab: params}
	\begin{tabular}{lll} 
		\hline
		Parameter & value (error) & units\\
		\hline
        $P_\mathrm{orb} $& $ 3.4939558(3) $& ${\rm h}$\\
        $P_\mathrm{spin}$ & $92.28(3)$ & ${\rm s}$ \\
        Spectral type & M$4.0\pm0.5$ & \\
        $ T_\mathrm{eff,2}$ & $3300(100)$ & K \\
        $ \log_\mathrm{10}(R_2/d)^2$ & -21.95(6) & \\
        $f(M)$ & 0.21(1) & $\rm M_\odot$ \\
        $ M_2^\textsuperscript{\textdagger}$ & $0.19-0.28$& $\rm M_\odot$\\
        $ d^\textsuperscript{\textdagger}$ & $1.25 (2)$& $\rm kpc$\\
        \hline
	\end{tabular}
\end{table}

\subsection{Constrains from Roche geometry} \label{sec: roche}

\begin{figure*}
      \centering
      \includegraphics[width=0.7\textwidth]{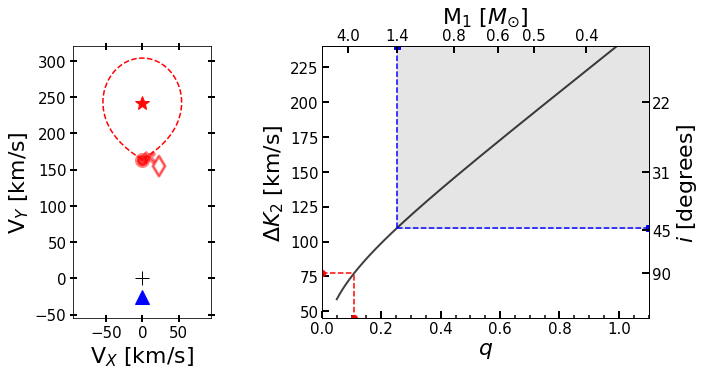}
    \caption[Constrains from Roche lobe geometry. ]{\textbf{ Left:} projected velocities in the orbital plane $\rm V_x$, $\rm V_y$ for the lines derived from our {GMOS} spectroscopy. The star indicates the centre of mass of the M-dwarf as estimated from the Na {\sc i} radial velocity curve. H$_\mathrm{\rm \alpha}$, He\,{\sc i}\,$\lambda 6678$ and the Ca\,{\sc ii} triplet are indicated with a red cross, a diamond and a circle respectively. The black cross and blue triangle are the centre of mass of the system and the WD respectively. The dashed red line is the Roche lobe of the companion for a mass ratio $\rm q=0.11$. While H$_\mathrm{\rm \alpha}$, and Ca\,{\sc ii} have a broadly consistent position in the side of the companion facing the compact object, the He\,{\sc i} line (open diamond), seems to be lying in the trailing face of the companion, probably owing to poor S/N at some specific orbital phases. This line was not included to determine the limit in the mass ratio $q_\mathrm{min}$. 
    \textbf{ Right:} The solid black line indicates the radial velocity difference ($\Delta K_2$), between the irradiated face and the centre of mass of the secondary as a function of the mass ratio $\rm q$ for a Roche lobe filling companion star. The left $y$-axis is the maximum $\Delta K_2$ for a given mass ratio. The right $y$-axis represents the required inclination to explain the observed $\Delta K_2$ given by $K_\mathrm{\rm Na\,\textsc{i}} - K_\mathrm{\rm Ca\,\textsc{iii}}$. The top $x$-axis represents the mass of the primary if we assume that the companion has the maximum density possible for this orbital period. The red dashed lines indicate the mass ratio for a system with $i=90{\rm ^o}$ with the observed $\Delta K_2$ corresponding to $\rm q = 0.11$. The blue dashed lines indicate the maximum inclination allowed for $ M_1= 1.4 \Msun$ and $\rm M_2=M_2^\mathrm{max}$, leading to the constrains $i \lesssim 45{\rm ^o}$ and $q \gtrsim 0.25$ indicated by the shaded area.}
    \label{fig: roche}
\end{figure*}

The radial velocities of the lines observed in \ref{sec: spectroscopy} allow us to obtain further information about the properties of the system. 
For an irradiated companion filling its Roche lobe in a close binary, the narrow emission lines typically originate in the irradiated face lying very close to the inner Lagrangian point. 
Therefore, the amplitude of their observed radial velocity curves ($K' $) is modulated by the projected geometry of the secondary's Roche lobe, i.e. the ``true'' semi-amplitude tracing the centre of mass of the secondary $K = K(K',q,i)$, is a function of the mass ratio of the system $q$, and the inclination of the system which affects the geometrical projection towards the observer. The effect on the mass ratio is known as the $K-$correction \citep[e.g.][]{K-Correction_MD2005ApJ...635..502M}, and it can be a very powerful tool to constrain fundamental parameters of binary systems. 
In the case of \src\ we also detect narrow absorption lines originating in the cold atmosphere of the companion star. These lines, namely Na\,{\sc i}, are a good tracer of the centre of mass of the secondary star \citep[e.g. ][but also see \citealt{Friend_NaI_shift:1990MNRAS.246..654F}]{Marsh2016Natur.537..374M}
. The effect of the $K-$correction is already noticeable in Fig.\,\ref{fig: folded rvs}, here we can see how all the lines share the same systemic velocity ($\gamma$) and orbital period, however the amplitude of the absorption line ($K_\mathrm{\rm Na\,\textsc{i}}$) is much larger than the one from the measured emission lines.

As we show in the left panel from Fig.\,\ref{fig: roche}, the geometrical effect induced by the $K-$correction can be easily visualized by projecting the velocities studied in Sec. \ref{sec: spectroscopy} into the $V_x$,$V_y$ plane\footnote{Where $V_x$ and $V_y$ were obtained from the radial velocity expressed as $V=\gamma - V_x\cos{(2\pi\phi_\mathrm{orb})} + V_y\sin{(2\pi\phi_\mathrm{orb}}$)}. In this panel, the emission lines (big red symbols) lie close to the projected Roche lobe surface of the companion star of a system seen edge-on (red dashed line), i.e. a system with inclination $i=90^{\rm o}$.
Provided that the companion, and therefore the emission line region, has to be contained within its Roche lobe, the minimum mass ratio required to engulf the emission lines sets a lower limit for the mass ratio. In the case of \src\ we find that the projected velocities of H$_\mathrm{\rm \alpha}$, and Ca\,{\sc ii} have a broadly consistent position in the irradiated face of the companion, while the He\,{\sc i} line lies in the trailing face of the companion, likely owing to poor S/N at some specific orbital phases (c.f. Fig.\,\ref{fig: folded rvs}).
Ignoring the position of He\,{\sc i} in the diagram, we find that the mass ratio must be $q\gtrsim0.11$. This lower limit in the mass ratio ($q=0.11$), produces the Roche lobe shown by the red dashed line in the left panel of Fig.\,\ref{fig: roche}.

The black line in the right panel of Fig. \ref{fig: roche}, represents the $K-$correction ($\Delta K_2 = K_{\rm Na\,\textsc{i}} - K_\mathrm{\rm Ca\,\textsc{iii}}$) as a function of the mass ratio ($q$; left and bottom axes, respectively), for a Roche lobe filling companion star as seen at $90^{\rm o}$ degrees. 
Using this lower limit in the mass ratio we can recast the mass function (eq. \ref{eq: f(m)}) as $M_1 = f(m)\,(1+q)^2/\sin^3{i}$, to obtain a lower limit on the primary and secondary masses, yielding $M_1>0.26\,\Msun$ and $M_2>0.028\,\Msun$.

The orbital period of the system also allow us to set an upper limit for the mass of the secondary. For a Roche lobe filling companion star, the maximum mean density it can have in $\mathrm{g\,cm}^{-3}$ is 
$\rho_2^\mathrm{max}=(0.43/{P_\mathrm{orb}})^2$, with ${\rm P_\mathrm{orb}}$ in days \citep{Eggleton1983ApJ...268..368E}. 
Assuming the semi-empirical mass-radius relation from \citet{Brown2022MNRAS.513.3050B} we obtain an upper limit for the mass of the secondary of $M_2^\mathrm{max}= 0.36\,{\rm M_\odot}$. 
The \citet{Brown2022MNRAS.513.3050B} relation should correct the over-sized problem of companion stars in close binaries \citep[e.g.][]{Knigge+2011ApJS..194...28K}. Therefore, we consider $M_2^\mathrm{max}$ as (or close to) the face value for mass of the companion star. 
With this assumption we can obtain a family of solutions compatible with the observed $\Delta K_2$ as a function of the inclination and mass ratio. The former is the solid black line shown in the right panel of Fig.\,\ref{fig: roche}, the corresponding inclination is shown in the right $y$-axis of this figure, the top $x$-axis in this figure shows the corresponding primary mass that would explain the observed $\Delta K_2$ provided $M_2 = M_2^\mathrm{max}$.
This allows us to set a tighter constraint on the minimum mass ratio of the system that would be compatible with the observed $K$-correction and having a WD as the compact object (see Sec.\,\ref{sec: wdp discussion} for a discussion on the nature of the compact object). In turn, the Chandrasekar limit will then provide a lower limit for the inclination of the system compatible with the observations, $i\lesssim45^{\rm o}$, leading to the shaded region in this figure indicating the allowed region of the parameter space. This inclination is in agreement with the maximum inclination required for the system to be a non-eclipsing binary. 

In short, the constrains on $M_2$ set by the orbital period of the system combined with the maximum $K-$correction required to explain the observed $\Delta K$ (assuming a Roche lobe filling donor), sets the maximum inclination and mass ratio allowed for a primary mass $M_1\leq1.4\,\Msun$ to be $i\lesssim 45^{\rm o}$ and $q\gtrsim0.25$ respectively.

\subsection{Distance to \src} \label{sec: distance}

\src\ is detected in {\it Gaia} Data Release\,3 (DR3) with the ID: DR3 2842721961392797312 \citep{GAIA_DR3_2023A&A...674A...1G}. Unfortunately, the parallax is poorly constrained ($\pi \simeq 0.9\pm0.5\ {\rm m.a.s.}$). However, the target is listed in the \citet{Bailer-Jones_GDR3:2021AJ....161..147B} catalogue, with an estimated (geometric) distance of $d\simeq1.5^{+1.5}_{-0.5}\,{\rm kpc}$.
We can further constrain the distance by combining the estimations of the distance modulus obtained in Sec.\,\ref{sec: mass function and companion} with the mass-radius relation from \citet{Brown2022MNRAS.513.3050B}. 
We find that for a typical configuration of Roche lobe filling semi-detached binaries within the observed range of parameters \citep[ $ 0.25\,\Msun< M_2 < 0.35\,\Msun$ and $P_\mathrm{orb}\approx3.5\,h$; ][]{Knigge+2011ApJS..194...28K} we find $1.3{\rm\,kpc} < d < 2.0{\rm\,kpc}$. 
In fact, if the mass configuration of \src\ is similar to \wdp\ \citep[i.e. $M_2\simeq 0.25\pm 0.5\,\Msun$;][]{Pelisoli_J1912_UV:2024MNRAS.527.3826P}, we obtain $1.1 {\rm\,kpc} < d < 1.7\,{\rm kpc}$, which is well in line with the distance from \citet{Bailer-Jones_GDR3:2021AJ....161..147B}. Given the similarities of the observational properties of \src\ with \arsco\ (Sec.\,\ref{sec: wdp discussion}) we can assume both sources have the same luminosity and make a direct comparison using the average \textit{Gaia} magnitude to obtain an estimate of the distance. Adopting the value from \cite{GAIA_DR3_2023A&A...674A...1G} we have a distance of $d = 117\pm0.5\,{\rm pc}$ and $G_\mathrm{\arsco}= 14.98(3)\,{\rm mag}$ for \arsco, and $G_\mathrm{J2306}= 19.9(2)\,{\rm mag}$ for \src. This yields a distance of $d = 1.25(2)\, {\rm kpc}$. Combining this value with the constrains from above, we obtain $0.19\,\Msun\lesssim M_2\lesssim 0.28\,\Msun$ for Roche lobe filling factors $\geq 0.8$, which is very similar to the constrains for \wdp. 

\begin{figure}
      \centering
      \includegraphics[width=0.5\textwidth]{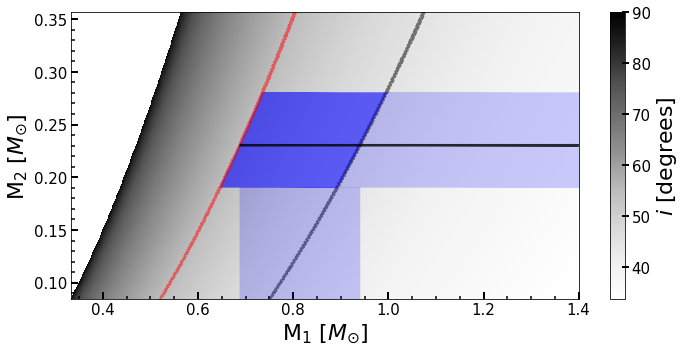}
    \caption{Mass of the primary vs secondary; the grey gradient is the corresponding inclination for each combination of masses determined from the mass function obtained in Sec.\,\ref{sec: mass function and companion}. The curves represent the maximum inclination for the system to be eclipsing ($i\lesssim55^{\rm o}$, red curve on the left side) and the maximum inclination allowed for $\rm M_2 = M_2^\mathrm{max}$ (black curve). The horizontal light blue region is the allowed secondary mass range compatible with our observations. The horizontal black line is the mass derived from the spectroscopic contribution of the donor at $\rm 1.25\,kpc$ which fills $\sim 90$ per cent of its Roche lobe (see Sec.\,\ref{sec: distance} for a discussion on the distance). The intersection of this mass range and the highlighted inclinations defines the vertical light blue region. This latter region is almost centred on the canonical mass of interacting WD binaries \citep[e.g. ][]{zorotovic2011,Pala2017:MNRAS.466.2855P}. A hard limit of $ M_1\gtrsim0.65\,\Msun$ can be set from this relation.
    }
    \label{fig: m1-m2 vs i}
\end{figure}

In figure\,\ref{fig: m1-m2 vs i} we show the relation between the primary and secondary masses with the inclination derived from the mass function obtained in Sec.\,\ref{sec: mass function and companion}. If we impose the condition that \src\ is not an eclipsing system, we find a lower limit on the WD mass of $ M_1 \gtrsim 0.65\,\Msun$. Interestingly, the intersection of the mass corresponding to a donor filling 90 per cent of its Roche lobe is centred around the canonical mass of WDs in interacting binaries \citep[i.e. $M_{\rm WD}\approx 0.8 \Msun$ e.g. ][]{zorotovic2011,Pala2017:MNRAS.466.2855P}, suggesting that the inclination of \src\ is relatively high ($i\sim 50^{\rm o}$).

\section{Discussion} \label{sec: discussion}

\subsection{The nature of the compact object in \src: defining the observational properties of WD pulsars} \label{sec: wdp discussion}

\begin{table*}
 \setlength\extrarowheight{2pt}
 \centering
 \captionof{table}{Summary table of White Dwarf Pulsars properties} \label{tab: WDp properties} 
 \label{my-label}
 \begin{tabular}{cc|c|c|c|c|c|}
  \cline{3-7}
  \multirow{2}{*}{}                            & \multirow{2}{*}{} & \multicolumn{5}{c|}{Observational properties of WD pulsars}   \\ \cline{3-7} 
                                         &                   &  Heated M-Dwarf\,$^{\dagger}$ & No steady accretion\,$^\ddagger$  & Pulsed emission\,$^\star$  &  $\uparrow$ Optical Polarization\,$^{\mathlarger{\ast}}$ & $\uparrow$ $\rm P_\mathrm{spin}/P_\mathrm{orb}$  \,$^\S$   
   \\\cline{5-5}
                                         &                   &   &   & X-ray~\vline~OPT/NIR~\vline~Radio  &   &   \\ \hline
  \multicolumn{1}{|c|}{\multirow{4}{*}{\rotatebox{90}{~~~Target}}} & \arsco                  & \checkmark  & \checkmark &\centering \checkmark ~~~~ \vline~~~~~~~\checkmark ~~~~~~~\vline ~~~~~~\checkmark&  \checkmark &  \checkmark  \\ \cline{2-7} 
  \multicolumn{1}{|c|}{}                       & \wdp      &  \checkmark & \checkmark &\centering \checkmark ~~~~ \vline~~~~~~~\checkmark ~~~~~~~\vline ~~~~~~\checkmark &  \checkmark & \checkmark \\ \cline{2-7} 
  \multicolumn{1}{|c|}{}                       & {J2306}                 & \checkmark & \checkmark &\centering $?$ ~~~~ \vline~~~~~~~\checkmark ~~~~~~~\vline ~~~~\checkmark$?$& $?$ & \checkmark  \\ \cline{2-7} 
  \hline
 \end{tabular}\par
 \smallskip
 \RaggedRight 
 $^\dagger$ The large amplitude modulation ($\sim 1.5\,\rm mag$) observed in the optical ligtcurves exhibiting one maximum per orbital cycle suggest the presence of a hot object. However, their spectrum does not show signatures of a hot WD, contrary to post-common envelope or sdB WD + M-dwarf  binaries \citep[e.g.][]{Parsons_NNSer:2010MNRAS.402.2591P, WD+BD_LCs:2023A&A...673A..90S}.\\
 $^\ddagger$ The low X-ray luminosity, absence of flickering in the light curves, extremely narrow emission lines, and relatively cool white dwarf all indicate a lack of persistent accretion.\\
 $^\star$ Pulsed non-thermal emission across the electromagnetic spectrum is one of the characteristic of WD pulsars. The amplitude of these pulses is modulated with the orbital period\\
 $^\mathlarger{\ast}$  In \arsco\ and \wdp, the pulsed emission exhibit significant polarization degree in at optical wavelengths. Reaching up to $\sim 40$ per cent polarisation degree and a polarisation fraction of the pulsed emission near to 90 per cent in \arsco\ \citep[][]{Buckley_ARSco:2017NatAs...1E..29B, Potter_Buckley_ARSco:2018MNRAS.481.2384P}. \\
 $^\S$ In general WD pulsars have $P_\mathrm{spin}/P_\mathrm{orb}\sim10^2$, which is larger than the typical observed in IPs, but less than the magnetic WD-propellers (see Fig.\,\ref{fig: porb-pspin}). 
\\ 
\end{table*}

\begin{figure}
    \centering
    \includegraphics[width=0.98\linewidth]{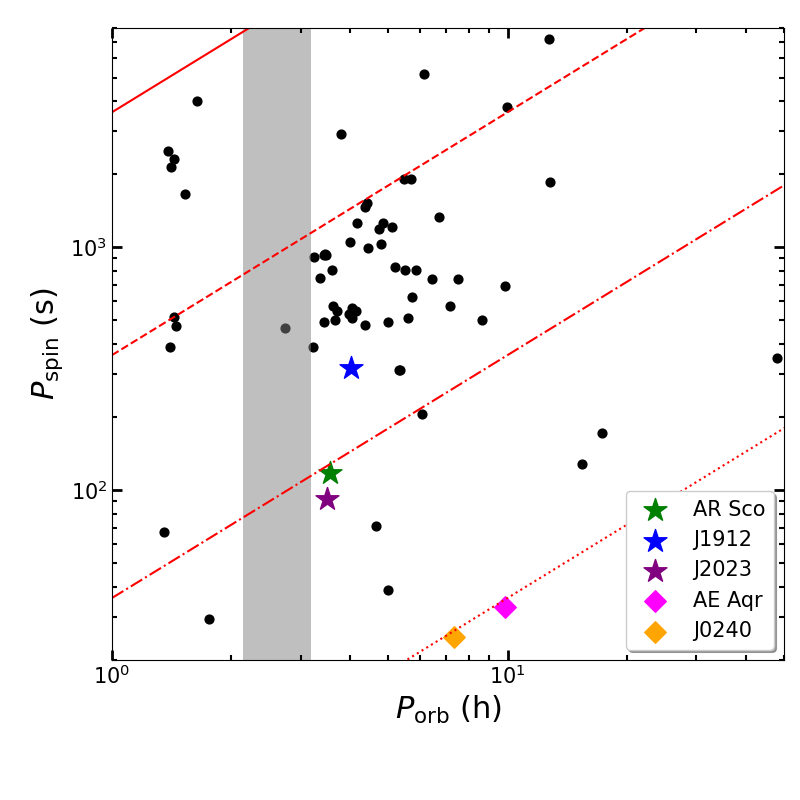}
    \caption{
    Spin period vs orbital period of confirmed IPs (back dots). Known WD pulsars and propeller systems are represented with coloured stars and diamonds respectively. The vertical gray area indicates the period gap, where there is a dearth of systems, as defined in \citet{Knigge+2011ApJS..194...28K}. 
    Diagonal lines correspond to spin periods of 1, 10, 100 and 1000 times the orbital period (solid, dashed, dash-dot and dotted line respectively). The most extreme systems are the propeller {AE Aqr} and {LAMOST J024048.51+195226.9} with $P_\mathrm{spin}/P_\mathrm{orb}\simeq10^3$. 
    The WD pulsars \arsco~ and \src~ are clustered around  $P_\mathrm{spin}/P_\mathrm{orb}\simeq10^2$, while \wdp~ lies much closer to the bulk of the IP population. Adapted from \citet{Mukai2017PASP..129f2001M}. 
    }
    \label{fig: porb-pspin}
\end{figure}

In section \ref{sec:analysis} we clearly identify \src\ as a spectroscopic binary with a $T\approx3300\,{\rm K}$ M-dwarf companion harbouring a rapidly spinning compact object. 
But, what is the nature of the compact object?
Given all the similarities with the known WD pulsars exposed throughout this research, the most natural interpretation is that the compact object is a WD where the short-term pulsations seen at $P_\omega$ are produced by the interaction of its magnetic field with the companion star's own field. 
In fact, the pulsed emission profile shown in Fig.\,\ref{fig: folded spin} is very similar to that of \arsco\ \citep[see fig 2. from][]{Marsh2016Natur.537..374M}, with the spin (or beat) period almost identical to the WD pulsar prototype. If the pulsed emission is the actual spin period of \src, or even the beat period between the orbit and the spin (which would be about 10 per cent larger than the spin period), the rotation frequency of the compact object in \src\ would be at least a factor of ten slower than the typical periods of magnetars, which are the slowest-rotating type of observed neutron stars \citep[e.g.][]{Rea2024ApJ...961..214R}. Even if we consider \src\ as a magnetar, it should be a very young system, a scenario that does not match with the (likely) old M-dwarf companion \citep[][]{Beniamini2019MNRAS.487.1426B}, and the low X-ray luminosity implied by the non-detection by {\it ROSAT} (see below). Therefore, we can rule out the possibility of \src\ harbouring a neutron star. 
In the remainder of this section we will summarize all the similarities with \arsco, \wdp\ and \src, and identify the common properties that define WD pulsars. All these properties are summarized in Table \ref{tab: WDp properties}.

One of the most distinctive observational properties of WD pulsars arises from their orbital modulation in the optical and the presence of minute-timescale pulses in their light curves. 
The short timescales are dominated by pulsed emission (see e.g. Fig.\,\ref{fig: folded spin}), contrary to accreting systems harbouring a magnetic WD (i.e. polars and intermediate polars) and there is no evidence of stochastic variability \citep[a.k.a. flickering;][]{Scaringi_rms-flux:2015SciA....1E0686S} signalling the lack of accretion. This pulsed emission is highly coherent and its amplitude is modulated with the orbital period, exhibiting maximum dispersion between orbital phases $\phi_\mathrm{\rm orb} \sim0.25-0.75$, associating the visibility of the emission region with the trailing inner face of the companion (i.e. L1). As shown in Fig.\,\ref{fig: porb-pspin}, the compact object of WD pulsars is spinning quite rapidly, ${P_\mathrm{spin} \sim 10^2 \times P_\mathrm{orb}}$. This is significantly faster than the bulk of the intermediate polars, but still a factor of ten slower than the known WD propellers \citep[i.e. {AE Aqr} and {LAMOST J024048.51+195226.9};][ i.e. systems where mass transferred from the companion is accelerated and flung away from the system by the magnetic field of the fast-spinning WD]{Schenker_AE_Aqr_2002MNRAS.337.1105S,Pelisoli_propeller:2022MNRAS.509L..31P}. In addition to this the fast rotating IPs ($P_\mathrm{spin}\lesssim 10^2$\,s) with orbital periods $P_\mathrm{orb}\lesssim 10$\,h are classified as low luminosity systems \citep[e.g. ][]{Joshi_IPs:2025AJ....169..269J}, perhaps suggesting a common evolutionary scenario for rapidly spinning interacting WD binaries. 

On longer timescales (hours) WD pulsar's light curves exhibit a ${\rm \sim 1.5\,mag}$ modulation with a single maximum per orbit, this is typically observed in detached systems harbouring a hot stellar remnant, where the surface of the companion star is heated by the strong radiation field from the primary \citep[e.g.][]{NNSer_UCAM_Brinkworth:2006MNRAS.365..287B}. 
However, the maximum of the light curve is not aligned either with quadrature or conjunction of the system as revealed by the spectroscopic orbital solution. This suggests a non-axi-symmetric origin of the heating mechanism, in line with the modulation of the pulsed emission. 
In addition to this, one would expect to observe the hot compact object heating the surface of the companion star, however, the blue part of the optical spectrum is clearly dominated by strong narrow emission lines superimposed on a featureless blue continuum (Fig. \ref{fig: full_spectrum}), implying a dominant non-thermal emission at short wavelengths. 
In this region, the Balmer emission series are the dominant lines observed, with a weak contribution of He\,{\sc i}\,$\lambda6678$\,\AA\ and even weaker He\,{\sc ii}\,$\lambda4686$\,\AA. In \src\ there is no actual evidence of He\,{\sc ii}\,$\lambda4686$\,\AA, which is particularly strong among accreting systems with rapidly spinning compact objects \citep[e.g.][]{Voikhanskaya_IPs_spectra:1987SvAL...13..250V}. All the core emission lines observed in WD pulsars are very narrow, to the point that we cannot resolve them in our Gemini/GMOS data for \src, implying in our case a $\rm FWHM\lesssim 70\,km\,s^{-1}$, another piece of evidence signalling the absence of mass transfer from the low mass companion onto the primary. 
The red end of the spectrum is dominated by molecular bands arising from the atmosphere of a relatively cold M-dwarf companion. At these wavelengths, the presence of the Ca triplet in emission reveals a temperature inversion in the atmosphere of the companion owing to the heating of its atmosphere. This heating is thought to be caused by the interaction of the magnetic field with the atmosphere of the M-dwarf \citep[e.g. ][]{Geng_ARSco:2016ApJ...831L..10G,Katz_ARSco_reconnection:2017ApJ...835..150K,Garnavich_ARSco_time_resolved_sp:2019ApJ...872...67G}. 
In addition to the narrow emission components, in \src\ we can identify (at least in some of the strongest lines), a broad- low-amplitude secondary emission component seen at least in H$\alpha$, 
similar to \arsco\ and \wdp. 
As can be seen in Fig.\,\ref{fig: trailed}, the strength of the emission lines in \src\ is also modulated with the orbital period. 
As discussed above, the same is true for the pulse fraction of the signal producing $P_\omega$, exhibiting a minimum roughly when the trailing face of the companion star is facing away from Earth. 
This may be related to the partial occultation of the high excitation region site likely located on the trailing face of the companion. A similar effect has been reported in \arsco\ by \citet{Garnavich_ARSco_time_resolved_sp:2019ApJ...872...67G}; these authors show how this region produces the bulk of He\,{\sc ii} $\lambda 4686$\,\AA~  in the WD pulsar prototype.

In general, the pulsed emission on WD pulsars is thought to arise predominately from synchrotron radiation, which dominates across the electromagnetic spectrum.. Since there is virtually no accretion in these systems, the X-ray emission is also dominated by the pulsed emission, at least in \arsco\ and \wdp. Unfortunately, we do not have any pointed X-ray observations of \src\ and it is not detected by the {\it ROSAT} all-sky survey \citep[][]{Voges1999A&A...349..389V}. However, we can use the upper limit of $\approx 0.004\,{\rm ct/s}$ from {\it ROSAT} to obtain an upper limit of $L_x<2\times 10^{-31} (d/1.2\,{\rm kpc})^2\,{\rm erg/s}$ in the $0.1-2.4\,{\rm keV}$ band assuming a simple power law with photon index of 2 and adopting the extinction to hydrogen column density conversion from \citet{Tolga2009_NH_to_EB-V}. This upper limit lies in the lower band of X-ray luminosities observed accreting WDs suggesting the actual luminosity for \src\ is lower than the typical CV \citep[e.g.][]{PretoriusKnigge2012MNRAS.419.1442P, PretoriusMukai2014MNRAS.442.2580P,Schwope2018A&A...619A..62S}. Combining this with the average SDSS $g$-band magnitude ($g\approx 21$\,mag) we obtain  $\log(L_X/L_{\rm opt})<0.21$, in line with other WD pulsars. On the other end of the electromagnetic spectrum, preliminary results from Sahu et al. (in prep) show a detection of \src\ with the Karl G. Jansky Very Large Array telescope (VLA) at 6\,cm (Band C), however, further analysis is required to determine the presence of pulsed emission similar to the one seen in other WD pulsars \citep[][]{AR_Sco_Stanway:2018A&A...611A..66S,Pelisoli2023NatAs...7..931P}.

\begin{figure}
      \centering
      \includegraphics[width=0.5\textwidth]{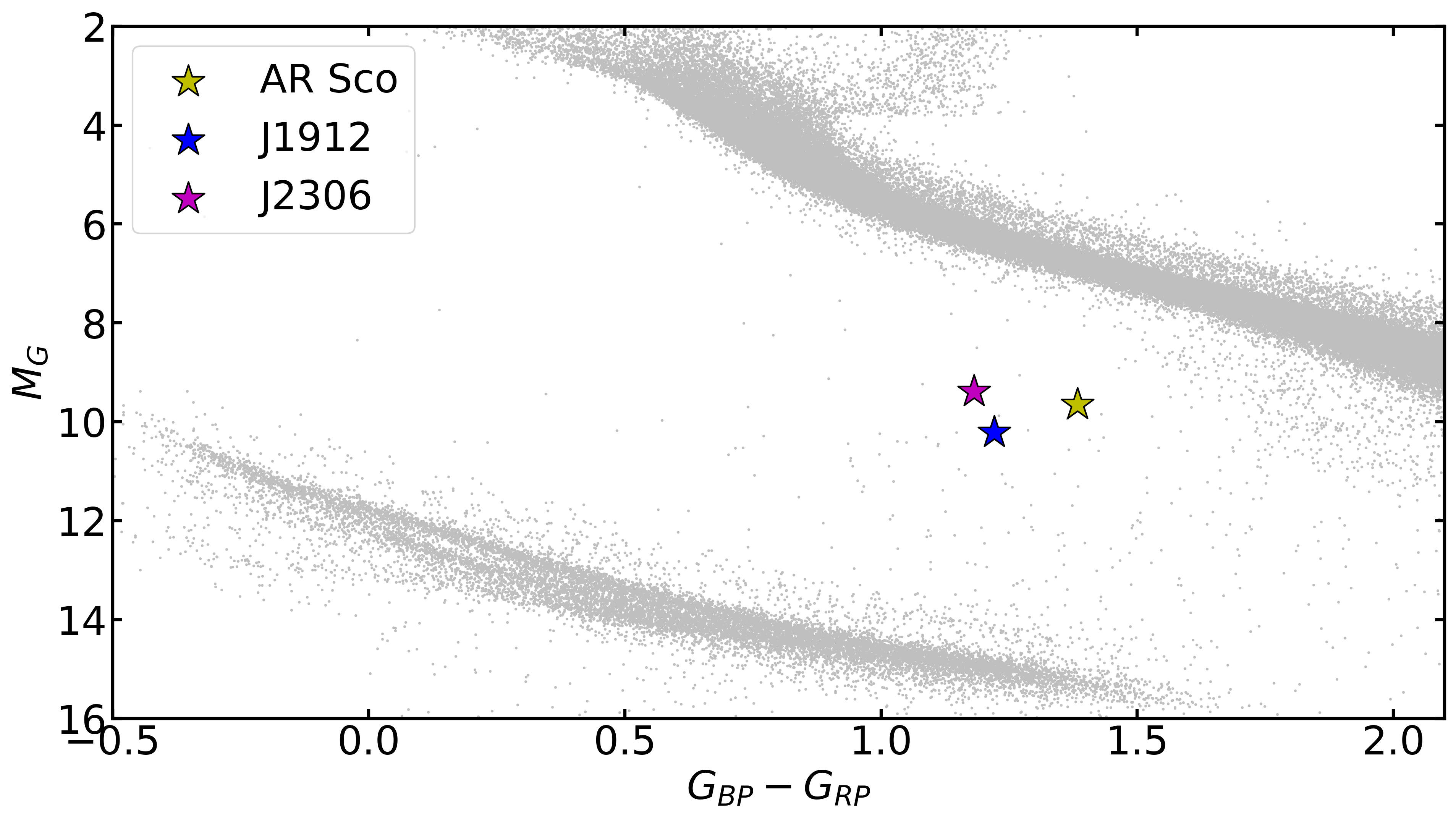}
    \caption{Gaia colour-magnitude diagram. \arsco, \wdp\ and \src\ are indicated with green, blue and pink stars respectively. All three systems are in a similar region of the parameter space between the WD sequence and the main sequence (bottom and top grey clusters, respectively).
    }
    \label{fig: CMD}
\end{figure}

Another of the common characteristics of WD pulsars is the strong linear polarization observed at optical wavelengths. \arsco\ is the most extreme case, exhibiting up to 40 per cent linear polarization \citep[][]{Buckley_ARSco:2017NatAs...1E..29B}. In \wdp\ the polarized emission is more modest, with up to 12 per cent of linear polarization, but with a pulse morphology and orbital phasing very similar to \arsco\ \citep[][]{Potter_Buckley_ARSco:2018MNRAS.481.2384P,Pelisoli2023NatAs...7..931P}. There are no photopolarimetric observations of \src, however, it is an excellent candidate given how similar this new WD pulsar is to \arsco. Finally, as we show in Fig.\,\ref{fig: CMD}, despite the small number of identified systems in the sample, the position of WD pulsars on the HR diagram seems to be clustered around absolute magnitudes $M_G\simeq10\,{\rm mag}$ in the {\it Gaia} $G$-band and $G_\mathrm{BP}-G_\mathrm{RP}\simeq1.25$ (see Sec. \ref{sec: distance} for a discussion on the distance), a relatively unexplored region of the {\it Gaia} colour-magnitude diagram \citep[e.g.] []{Abrahams_CVs_GAIA:2022ApJ...938...46A,Pelisoli_Gaia_search:2025MNRAS.tmp..722P}. This suggests that the (time averaged) spectral components dominating at these wavelengths may be produced by the same physical process. 
The low number of WD pulsar discovered, combined with their clustering in the parameter spaces discussed in this section, suggests that the formation of WD pulsars requires fairly constrained stellar and orbital properties in their progenitor systems. This interpretation aligns with previous theoretical studies by \citet{Schreiber_B_in_WDs:2021NatAs...5..648S}.

In summary, the observational properties of \src\ presented in this research (i.e. spin and orbital periods, pulse morphology, companion star) reveal this new system as a sibling of \arsco\ and the third system of the class.

\subsection{Evolutionary pathways}

The evolution of WD pulsars is an interesting open topic, in particular in connection with long period radio transients (LPTs). The origin of LPTs is uncertain, but two were found to be binary systems harbouring a WD with an M-dwarf companion \citep[][]{Hurley-Walker_LPT:2024ApJ...976L..21H, deRuiter_LPT:2025NatAs.tmp...78D} which exhibit relatively bright radio pulses with a frequency consistent with the orbital period \citep[][]{deRuiter_LPT:2025NatAs.tmp...78D, Rodriguez_LPT:2025A&A...695L...8R}. Given the observational properties of this new class of radio transients, in particular the lack of a detected WD spin period, they have been postulated as an intermediate evolutionary configuration between WD pulsars and strongly magnetized CVs, i.e. polars, where the WD spin has synchronised with the orbit. 
The low temperatures observed in the WDs in LPTs ($5-7\times 10^3\,\rm K$) require the hypothetical progenitors (WD pulsars) to not have had any accretion-induced heating episode in the past; in fact, their temperature must have been already relatively low during the WD pulsar stage \citep[c.f.][]{Townsley2004ApJ...600..390T, Althaus_WDevo:2010A&ARv..18..471A}. 
However, the observed WD temperature of the closest known WD pulsars \citep[$T_\mathrm{WD}\simeq11500\,{\rm K}$;][]{Garnavich_ARSco_UV:2021ApJ...908..195G,Pelisoli_J1912_UV:2024MNRAS.527.3826P} challenges this hypothesis. In contrast, this high temperature is consistent with WD pulsars having undergone an accretion-driven heating (and consequent spin up) episode in the past. 
This is consistent with the evolutionary model of \citet{Schreiber_B_in_WDs:2021NatAs...5..648S} and also with the pulsed emission mechanism proposed by \citet{Lyutikov_WDp_emission:2020arXiv200411474L}, which additionally requires the WD to have a low magnetic field, in agreement with recent observational constrains of AR\,Sco's WD \citep{Barrett_ARSco_SED:2025arXiv250506468B}.

A potential evolutionary connection has also been drawn with magnetic propellers, which are proposed to share the same emission model as WD pulsars in \citet{Lyutikov_WDp_emission:2020arXiv200411474L}. At a first glance this in agreement with the parameter space occupied by these two types of systems (see Sec.\,\ref{sec: wdp discussion}). 
However, there is a fundamental difference between them; {AE Aqr} is thought to have gone through a thermal timescale mass transfer \citep[TTSMT,][]{Paczynski1969ASSL...13..237P} as a consequence of a relatively massive donor at the time at which the system came into contact. This TTSMT, in turn, spun up the WD to match the present-day high spin periods \citep[c.f.][]{SchenkerKing2002ASPC..261..242S}. 
During this process, the companion star loses a significant part of its envelope, changing its external composition. The footprint of such a process can be detected in the strength of the ultraviolet spectral lines. As shown in \citet{CastroSegura2023}, the extreme line ratios seen in some accreting binaries \citep[e.g.][]{Mauche1997ApJ...477..832M,Haswell2002MNRAS.332..928H,Schenker_AE_Aqr_2002MNRAS.337.1105S, Gansicke2003ApJ...594..443G} can only be explained by the CNO-equilibrium abundances of a (previously) relatively massive companion star, suggesting that the companion stars have lost their hydrogen envelope. While ultraviolet spectroscopy of {AE Aqr} matches this scenario, the same cannot be said about the two WD pulsars observed in the UV \citep{Garnavich_ARSco_UV:2021ApJ...908..195G,Pelisoli_J1912_UV:2024MNRAS.527.3826P}, suggesting that WD pulsars and magnetic propellers have indeed followed a different formation channel. 

\section{Conclusion} \label{sec: conclusion}
We presented the discovery of a new WD pulsar, \srclong. We have used long-term photometric data from ZTF to determine a precise orbital period of $P_\mathrm{orb} = 3.4939558(3)$ h. Fast photometry obtained with ULTRASPEC was used to determine the spin period $ P_\mathrm{spin}= 92\,{\rm s}$. Folding the light curve onto the spin period reveals a double pulsed emission profile with major and minor peaks similar to the that observed in \arsco. The orbital ephemeris was obtained through time-resolved spectroscopy and we used the amplitude from the radial velocity solutions to determine the mass function of the system. \src's stellar companion was constrained to be a M$4.0\pm0.5$ M-dwarf with an effective temperature $T_\mathrm{eff}\approx 3300\,{\rm K}$. By subtracting the non-stellar emission we estimated the normalization factor ($R_2/d$). 
We estimated the distance to the system to be about $\rm 1.25\,kpc$, which in turn, constrains the companion to be in the mass range $ 0.19\,\Msun \lesssim M_2 \lesssim 0.28\,\Msun$ for Roche lobe filling factors $\ge 0.8$. All these properties reveal this new object to be a sibling of \arsco\ and the third system of the class of WD pulsars. 

We have compiled a comprehensive table of the observational properties exhibited by all three known WD pulsars, establishing a benchmark that will inform the classification of future discoveries within this rare class of objects.
We discussed the proposed evolutionary scenarios for WD pulsars. We ruled out a potential evolutionary connection with the recently discovered long-period radio transients (LPTs) hosting white dwarfs, due to the significantly higher temperatures observed in WD pulsars ($T_{\mathrm{WD}} \simeq 11500\,\mathrm{K}$) compared to those in LPTs ($5000$–$7000\,\mathrm{K}$). This temperature difference implies cooling timescales that are too long to support a direct evolutionary link.
Finally, we considered the possibility that the progenitor evolutionary channel of WD pulsars is the same as that of magnetic propellers (i.e., AE Aqr-like systems). If this were the case, WD pulsars would have been spun up during a thermal-timescale mass transfer (TTMT) phase, during which the system would be detectable as a super-soft X-ray source. Such an episode would remove most of the companion's envelope, thus exposing its core. However, there is no observational evidence of such a scenario in known WD pulsars. Therefore, we can rule out the possibility that WD pulsars were spun up via a TTMT phase. At the time of writing, the only theory that successfully explains the observed configuration of WD pulsars is the model proposed by \citet{Schreiber_B_in_WDs:2021NatAs...5..648S}, which requires the white dwarf to have been magnetized prior to the onset of accretion \citep[e.g.,][]{Isern_B_in_WDs:2017ApJ...836L..28I}. However, the origin of the magnetic field in WD pulsars remains a mystery.

\section*{Acknowledgements}

We dedicate this work to the memory of Prof. Tom Marsh, whose contributions to the development of ULTRASPEC and characterisation of \arsco, the first known white dwarf pulsar, have left a lasting legacy in the field.

NCS and DLC acknowledge support from the Science and Technology Facilities Council (STFC) grant ST/X001121/1. IP acknowledges support from the Royal Society through a University Research Fellowship (URF\textbackslash R1\textbackslash 231496). DS acknowledges support from the Science and Technology Facilities Council (STFC), grant numbers ST/T007184/1, ST/T003103/1, ST/T000406/1, ST/X001121/1 and ST/Z000165/1. This project has received funding from the European Research Council (ERC) under the European Union’s Horizon 2020 research and innovation programme (Grant agreement No. 101020057).  AA acknowledges support from Naresuan University (NU), and National Science Research and Innovation Fund (NSRF), grant no. R2568B019.

This work has made use of data obtained at the Thai National Observatory on Doi Inthanon, operated by NARIT.

This research made use of Astropy, a community-developed core Python package for Astronomy \citep{Astropy2018AJ....156..123A}. Based on observations obtained at the international Gemini Observatory, a program of NSF NOIRLab, which is managed by the Association of Universities for Research in Astronomy (AURA) under a cooperative agreement with the U.S. National Science Foundation on behalf of the Gemini Observatory partnership: the U.S. National Science Foundation (United States), National Research Council (Canada), Agencia Nacional de Investigaci\'{o}n y Desarrollo (Chile), Ministerio de Ciencia, Tecnolog\'{i}a e Innovaci\'{o}n (Argentina), Minist\'{e}rio da Ci\^{e}ncia, Tecnologia, Inova\c{c}\~{o}es e Comunica\c{c}\~{o}es (Brazil), and Korea Astronomy and Space Science Institute (Republic of Korea).

ZTF is supported by the National Science Foundation under Grants No. AST-1440341 and AST-2034437 and a collaboration including current partners Caltech, IPAC, the Oskar Klein Center at Stockholm University, the University of Maryland, University of California, Berkeley , the University of Wisconsin at Milwaukee, University of Warwick, Ruhr University, Cornell University, Northwestern University and Drexel University. Operations are conducted by COO, IPAC, and UW.

Funding for the Sloan Digital Sky Survey V has been provided by the Alfred P. Sloan Foundation, the Heising-Simons Foundation, the National Science Foundation, and the Participating Institutions. SDSS acknowledges support and resources from the Center for High-Performance Computing at the University of Utah. SDSS telescopes are located at Apache Point Observatory, funded by the Astrophysical Research Consortium and operated by New Mexico State University, and at Las Campanas Observatory, operated by the Carnegie Institution for Science. The SDSS web site is \url{www.sdss.org}.

SDSS is managed by the Astrophysical Research Consortium for the Participating Institutions of the SDSS Collaboration, including Caltech, The Carnegie Institution for Science, Chilean National Time Allocation Committee (CNTAC) ratified researchers, The Flatiron Institute, the Gotham Participation Group, Harvard University, Heidelberg University, The Johns Hopkins University, L'Ecole polytechnique f\'{e}d\'{e}rale de Lausanne (EPFL), Leibniz-Institut f\"{u}r Astrophysik Potsdam (AIP), Max-Planck-Institut f\"{u}r Astronomie (MPIA Heidelberg), Max-Planck-Institut f\"{u}r Extraterrestrische Physik (MPE), Nanjing University, National Astronomical Observatories of China (NAOC), New Mexico State University, The Ohio State University, Pennsylvania State University, Smithsonian Astrophysical Observatory, Space Telescope Science Institute (STScI), the Stellar Astrophysics Participation Group, Universidad Nacional Aut\'{o}noma de M\'{e}xico, University of Arizona, University of Colorado Boulder, University of Illinois at Urbana-Champaign, University of Toronto, University of Utah, University of Virginia, Yale University, and Yunnan University.

\section*{Affiliations}
\noindent
{\it
$^{1}$Department of Physics, University of Warwick, Gibbet Hill Road, Coventry, CV4 7AL, UK\\
$^{2}$Department of Physics, Faculty of Science, Naresuan University, Phitsanulok 65000, Thailand\\
$^{3}$Universidade Federal do Rio Grande do Sul (UFRGS), Departamento
de Astronomia - IF Av. Bento Gonçalves, 9500 Campus do Vale
91540-000 - Porto Alegre - RS, Brazil\\
$^{\sheffield}$Astrophysics Research Cluster, School of Mathematical and Physical Sciences, University of Sheffield, Sheffield, S3 7RH, United Kingdom\\
$^{\iac}$Instituto de Astrof\'{i}sica de Canarias, E-38205 La Laguna, Tenerife, Spain\\
$^{\cork}$School of Physics, University College Cork, Cork, T12 K8AF, Ireland\\
$^{\cork}$School of Physics, University College Cork, Cork, T12 K8AF, Ireland\\
$^{\oklahoma}$Homer L. Dodge Department of Physics and Astronomy, University of Oklahoma, 440 W. Brooks Street, Norman, OK 73019, USA
$^{\colorado}$JILA, University of Colorado and National Institute of Standards and Technology, 440 UCB, Boulder, CO 80309-0440, USA\\
$^{\hamburg}$Hamburger Sternwarte, University of Hamburg, Gojenbergsweg 112, 21029 Hamburg, Germany\\
$^{10}$South African Astronomical Observatory, P.O Box 9, Observatory, 7935 Cape Town, South Africa\\
$^{11}$Department of Astronomy, University of Cape Town, Private Bag X3, Rondebosch 7701, South Africa
}
\section*{Data Availability}

The data underlying this article is publicly available in the corresponding telescope archives: {\em Sloan Digital Sky Survey} \url{https://www.sdss.org}, {\em Zwicky Transient Facility} \url{https://www.ztf.caltech.edu} and {\em Gemini} \url{https://archive.gemini.edu}
Remaining data will be shared on reasonable request to the corresponding author.



\bibliographystyle{mnras}




\appendix

\section{Long-term photometric light curves}
\begin{figure}
      \centering
      \includegraphics[width=0.5\textwidth]{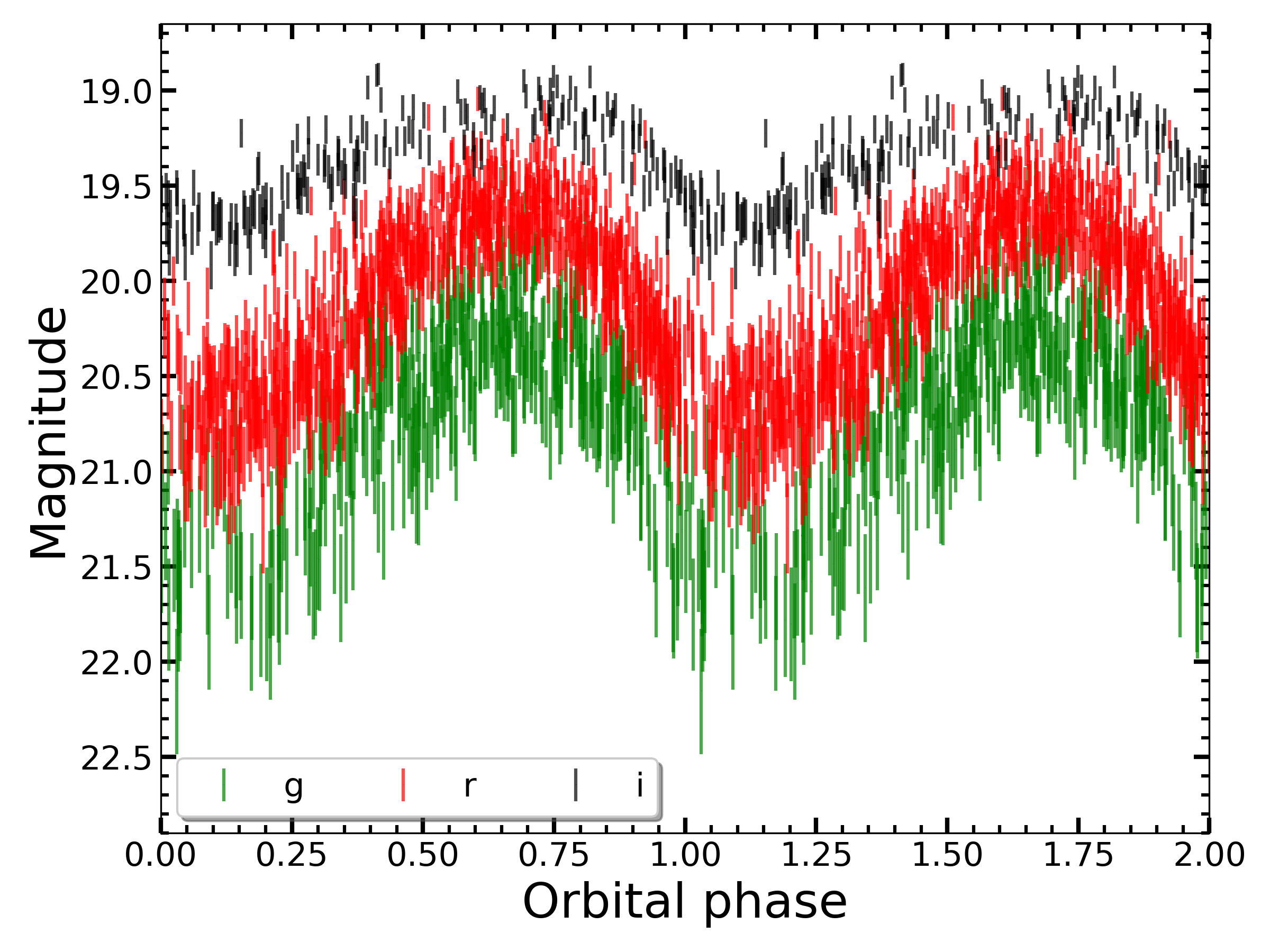}
    \caption{Light curves from ZTF folded into the orbital period.}
    \label{fig: folded lc}
\end{figure}

Photometric light curves from ZTF folded into the orbital period obtained from multi-band Lomb-Scargle analysis are shown in Fig.\,\ref{fig: folded lc}. The orbital phase was obtained from spectroscopic measurements presented in \ref{sec: spectroscopy}.



\bsp	
\label{lastpage}
\end{document}